\documentstyle[12pt,epsfig]{article}

\def\shiftdown#1{#1\llap{\lower.04ex\hbox{#1}}}

\begin{document}

\vspace*{0.3cm}

\begin{center}
{\bf {\large 
On relationship between conformal transformations and 
broken chiral symmetry. 
} }

\end{center}

\begin{center}
{\large {\it A.\ I.\ Machavariani$^{\diamond \ast}$ }}

{\em $^{\diamond }$ Joint\ Institute\ for\ Nuclear\ Research,\ Moscow
Region \\
141980 Dubna,\ Russia}\\[0pt]
{\em $^{*}$ High Energy Physics Institute of Tbilisi State University,
University Str. 9 }\\[0pt]
{\em 380086 Tbilisi, Georgia }
\end{center}

\vspace{0.5cm} \medskip

\begin{abstract}

Starting with the conformal transformations in the momentum space,
the nonlinear $\sigma$-model and the standard model
with the spontaneous broken $SU(2)\times U(1)$ symmetry
are reproduced. 
The corresponding chiral Lagrangians are given in the five dimensional form
because for the conformal transformations of the four-momentum $q_{\mu}$
( $q'_{\mu}=q_{\mu}+h_{\mu}$, $q'_{\mu}=\Lambda^{\nu}_{\mu}q_{\nu}$, 
$q'_{\mu}=\lambda q_{\mu}$ and  $q'_{\mu}=-M^2q_{\mu}/q^2$) 
the equivalence rotations in the 6D space
were used.
The derived five dimensional Lagrangians consists of the parts
defined in the two different region  $q_{\mu}q^{\mu}\pm q_5^2=\pm M^2$
which are connected by 
 the inversion $q'_{\mu}=-M^2 q_{\mu}/q^{2}$, where $M$
is a scale parameter.
For the $\sigma$-model 
$M$  is determined by the pion mass  $M^2=
m_{\pi}^2/2$. 
For the 5D Lagrangian with the spontaneous broken
 $SU(2)\times U(1)$ symmetry the scale parameter
$M^2$ is defined by the Higgs particle mass
$8m^2_{Higgs}=9M^2$.

Unlike to the usual four-dimensional formulation
in the present approach  
the chiral symmetry breaking terms are 
obtained from the conformal transformations and
it is demonstrated, that the corresponding
interaction parts of
Lagrangians have the opposite sign in the 
regions $q^{\mu}q_{\mu}>M^2$ and $0\le q^{\mu}q_{\mu}<M^2$.

\end{abstract}

\newpage

\begin{center}
{\bf{ Introduction}}
\end{center}

\vspace{0.5cm}

Conformal transformations in the configuration space 
consist of the complete set
of transformations  
$x'_{\mu}=x_{\mu}+a_{\mu}$, $x'_{\mu}=\Lambda^{\nu}_{\mu}x_{\nu}$, 
$x'_{\mu}=\lambda x_{\mu}$ and  $x'_{\mu}=
-\ell^2x_{\mu}/x^2$.
The conformal transformations  of four momentum 
$q_{\mu}$  
can be performed in the same manner as 
conformal transformations in the  coordinate space.
In the framework of the   
Dirac geometrical model \cite{Dir,Kas,Ca1,Ca2,M1,Gatto,Alf,Kon,BK}, 
each of the conformal transformations can be single-valued reproduced
via the appropriate 6D rotation with the invariance 6D form

$$\kappa_A\kappa^A\equiv\kappa_{\mu}\kappa^{\mu}+\kappa_5^2-\kappa_6^2=0,
\eqno(I.1)$$

where the four momentum
$q_{\mu}$ ($\mu=0,1,2,3$) is defined as
$q_{\mu}=M \kappa_{\mu}/{(\kappa_5+\kappa_6)}$ and
$M$ is a scale parameter.
The 6D cone $\kappa_A\kappa^A=0$ (I.1) or corresponding surface 

$$q_{\mu}q^{\mu}+M^2
{{\kappa_5-\kappa_6}\over{\kappa_5+\kappa_6}}=0
\eqno(I.2)$$
are invariable under any conformal transformation of $q_{\mu}$  
even when the conformal invariance is violated by the mass or other
dimensional parameters of considered particle.
In the present paper we use the invariance of 6D forms (I.1) or (I.2)
as a constraint for
the derivation of the equation of motion for a interacting 
massive field. In particular, we shall project conditions
$\kappa_A\kappa^A=0$  on the two  5D surface
$$q_{\mu}q^{\mu} + q_5^2= M^2\ \ \ and\ \ \ q_{\mu}q^{\mu} - q_5^2= -M^2,
\eqno(I.3)$$
so that these 5D hyperboloids 
remains to be also invariant under the conformal transformations. 
This enables us to introduce the constraints 
$\Bigl(q_{\mu}q^{\mu} \pm q_5^2 \mp M^2\Bigr)\Phi(q,q_5)=0$
for a 5D field operator $\Phi(q,q_5)$.
Afterwards we  shall construct corresponding  5D Lagrangians 
appropriate to the
5D equations of motion and consider their 4D reductions.

The 4D rotation and dilatation subgroups of the conformal group
in the coordinate and in the momentum space are
simply connected.
Therefore only
the four-momentum translation $q'_{\mu}=q_{\mu}+h_{\mu}$ and the 
four-momentum inversion $q'_{\mu}=-M^2q_{\mu}/q^2$
require a special attention.
In the homogeneous 4D momentum space invariance under the
translation $q'_{\mu}=q_{\mu}+h_{\mu}$ for the charged particle 
can be interpreted as invariance under  a gauge transformation
and for a  charged field operator we have
$\Phi'_{\gamma}(x)=\Phi_{\gamma}^h(x)=e^{ih_{\mu}x^{\mu}}\Phi_{\gamma}(x)$.
We shall show, that
for a neutral field the four-momentum translation
has  the more complicated
form $\Phi'_{\gamma}(x)=\Phi_{\gamma}^{h}(x)$ which
does not change the creation or annihilation operator
of the considered particle. 
Therefore the
scattering ${\cal S}$-matrix is invariant under translations in the
four-dimensional momentum space $q'_{\mu}=q_{\mu}+h_{\mu}$.

The numerous applications of the conformal transformations in the 
coordinate space are presented in  
many books and review papers (see for instance 
\cite{M1,Alf,Gatto,Kon,BK,BaR,BR,Braun,Fradkin,Todorov}).
Conformal transformations of the field operators 
and corresponding equation of motion 
in the momentum space were considered in ref.
\cite{Kas,Ca2,BR,Budinich}. In these papers  conformal
transformations were performed in the coordinate space 
and  followed relations were given in the momentum space 
using the Fourier transform.
Two particular features 
determine the special interest to the conformal transformations in
the momentum space\cite{BR}. First, the real observables of the
particle interactions, like the cross sections and the corresponding
scattering amplitudes, are determined in the momentum space.
Secondly, the accuracy of the measurement of the particle coordinates
is in principle restricted by the Compton length of this
particle. Moreover in the conformal invariant case
 determination of the coordinates
 of the massless particle 
generates an additional essential trouble 
(see  \cite{BaR} ch. 20 and \cite{Bacr}).

The 5D formulation of the quantum  field theory 
with the invariance form 
$q_{\mu}q^{\mu} + q_5^2= M^2$ or $q_{\mu}q^{\mu} - q_5^2= -M^2$ 
was suggested in refs.  \cite{K1,K2,K3},
where the scale  parameter $M$ was interpreted as the 
fundamental (maximal) mass
and its inverse $1/M$ as the fundamental (minimal) 
length \cite{Heisenberg,Markov}.
In the present formulation $M$ has the meaning of a auxiliary mass parameter
which indicates the scale of the conformal symmetry breaking.
In this paper we have  defined $M$ 
using the well known spontaneous  symmetry breaking models.
In particular, for the  nonlinear $\sigma$ model  $M$ is fixed via
the mass of $\pi$ meson and for the present 5D generalization of the
standard model with the spontaneous broken
 $SU(2)\times U(1)$ symmetry the scale parameter
$M$ is determined by the  mass of the Higgs particle.

 In the first section we consider a conformal transformations in the 
4D momentum space and we shall show, that
the scattering ${\cal S}$-matrix is invariant  
under the four-momentum translation.
Five dimensional projections in the momentum space 
with corresponding 5D and 4D  equations of motion
 are considered in the sections 2 and 3.
Sections 4, 5 and 6 are devoted to the 5D
Lagrangian approach. Interaction part of 5D Lagrangian
${\cal L}_{INT}(x,x_5)$ in the framework of the
 nonlinear sigma model is constructed in sect. 5
using the boundary condition for the fifth 
coordinate.  
5D theories with the gauge transformations
and a model of the chiral $SU(2)\times U(1)$ Lagrangian 
is presented in ch.6.
The summary and outlook in ch. 7 is devoted to the short 
comparison 
with the other 5D field-theoretical and
to the main features of the present formulation.

\begin{center}
{\bf{1. \ Conformal transformations 
 in the 4D momentum space and the scattering ${\cal S}$-matrix}}
\end{center}

\vspace{0.5cm}

Conformal group of a four-momentum 
 $q_{\mu}$ $(\mu=0,1,2,3)$  transformations consists of the following   ¨
transformations:\newline
translations
$$T(h):\ \ \ \ \ \ \ \ \ \ \ \
\ \ q_{\mu}\longrightarrow q_{\mu}'=q_{\mu}+h_{\mu},\eqno(1.1a)$$
rotations

$$R(\Lambda):\ \ \ \ \ \ \ \ \ \ \ \ \ \ \ \ \ q_{\mu}\longrightarrow
q_{\mu}'=\Lambda_{\mu}^{\nu}q_{\nu}
,\eqno(1.1b)$$
dilatation
$${\cal D}(\lambda):\ \ \ \ \ \ \ \ \ \ \ \ \ \ \ \ \ \ \ \
q_{\mu}\longrightarrow q_{\mu}'=e^{\lambda} \
q_{\mu},\eqno(1.1c)$$ 
and inversions
$$I(M^2):\ \ \ \ \ \ \ \ \ \ \ \ \ \ \ \ \ \ \ \
q_{\mu}\longrightarrow
q_{\mu}'=-M^2 q_{\mu}/q^2,\eqno(1.1d)$$ 
where a scale parameter $M$ insures the correct dimension of  $q_{\mu}'$.
Translation $T({\hbar})$ and inversions $I(M^2)$ form the special conformal
transformations
$${\cal K}(M^2,{\hbar})\equiv I(M^2)T({\hbar})I(M^2):\ \ \ \ \
q_{\mu}\longrightarrow q_{\mu}'={{ q_{\mu}-{\hbar}_{\mu} q^2/ M^2} \over
{1-2q_{\nu}{\hbar}^{\nu}/M^2+{\hbar}^2q^2/M^4} }.\eqno(1.1e)$$
\par
Obviously, $q_{\mu}$ in (1.1a)-(1.1e) is off mass shell 
($q_o\ne\sqrt{{\bf q}^2+m^2}$). Hereafter the on mass shell 
4D momenta will be denoted as $p_{\mu}$ ($p^2=m^2$, $p_o\ge 0$).

Following the Dirac geometrical model \cite{Dir},
transformations (1.1a)-(1.1e) may by realized as rotations in the 
6D space with the metric
$g_{AB}=diag(+1,-1,-1,-1,+1,-1)$ and on the 6D cone

$$\kappa^2\equiv\kappa_A\kappa^A=
\kappa_{\mu}\kappa^{\mu}+\kappa_5^2-\kappa_6^2=0,\eqno(1.2)$$
where according to  conformal covariant formulation \cite{M1,M2}, 

$$q_{\mu}={ {\kappa_{\mu}}\over{\kappa_{+} }};\ \ \ \
\kappa_{+}=(\kappa_{5}+\kappa_{6})/M;\ \ \ \
\mu=0,1,2,3;\eqno(1.3)$$ 
where $\kappa_{+}$ is a dimensionless scale parameter and
$\kappa_{\mu}$, $q_{\mu}$ and $M$ have the same dimensions
 in the system of units $\hbar=c=1$.


\par
{\underline {\bf Conformal transformations (1.1a)-(1.1e)
of a particle field operator 
$\Phi_{\gamma}(x)$:}}

The particle field operator $\Phi_{\gamma}(x)$ 
with a spin-isospin quantum numbers $\gamma$ is usually expanded in the
positive and in the negative frequency parts in the 3D Fock space
$$\Phi_{\gamma}(x)=\int {{d^3 p}\over{(2\pi)^3 2\omega_{{\bf p}}} }
\Bigl[
a_{{\bf p}{\gamma}}(x_0)e^{-ipx}+{b^+}_{{\bf p}{\gamma}}(x_0)e^{ipx} \Bigr];
\ \ \ p_o\equiv \omega_{{\bf p}}
=\sqrt{ {\bf p}^2+m^2},\eqno(1.4a)$$
where in the asymptotic regions
$a_{{\bf p}{\gamma}}(x_0)$ and ${b^+}_{{\bf p}{\gamma}}(x_0)$ 
transforms into particle
 (antiparticle) annihilation (creation) operators
$lim_{x_0\to\pm \infty}<m|a_{\bf p\gamma}(x_0)|n>= <m|a_{\bf
p\gamma}(out,in)|n>$; ¨ $lim_{x_0\to\pm \infty}<m|{b^+}_{\bf
p\gamma}(x_0)|n>= <m|{b^+}_{\bf p\gamma}(out,in)|n>$, where
$<n|,|m>$ denotes  some n,m-particle asymptotic states.
On the other hand, $\Phi_{\gamma}(x)$  may be expanded in the 4D momentum 
space

$$\Phi_{\gamma}(x)=\int {{d^4 q}\over{(2\pi)^4  } }
\Bigl[{\Phi_{\gamma}^{(+)}} (q)e^{-iqx}+
{\Phi_{\gamma}^{(-)}}^+ (q)e^{iqx}\Bigr].\eqno(1.4b)$$

After comparison of the  expressions (1.4a) and (1.4b) we get

$${{e^{-i\omega_{\bf p}x_o}}\over{2\omega_{\bf p}}}
a_{\bf p\gamma}(x_o)=
\int {{dq_o}\over{2\pi}} {\Phi_{\gamma}^{(+)}} (q_o,{\bf p}) e^{-iq_ox_o}
\eqno(1.5a)$$
$$=i{{e^{-i\omega_{\bf p}x_o}}\over{2\omega_{\bf p}}}
\sum_{\beta}\int d^3x
<0|\Phi_{\beta}(x)|{\bf p}\gamma> \Bigl[{{\partial
\Phi_{\beta}(x)} \over{\partial x_o}}-i\omega_{\bf p}
\Phi_{\beta}(x)\Bigr]\eqno(1.5b)$$ and
$${{e^{i\omega_{\bf p_a}x_o}}\over{ 2\omega_{\bf p_a}} }
{b^+}_{\bf p_a\gamma}(x_0)=
\int {{dq_o}\over{2\pi}} {\Phi_{\gamma}^{(-)}}^+(q_o,{\bf p_a})e^{iq_ox_o}
\eqno(1.6a)$$

$$=-i{{e^{i\omega_{\bf p_a}x_o}}\over{ 2\omega_{\bf p_a}} }
\sum_{\beta}\int d^3x
<{\bf p}_a\gamma|\Phi_{\beta}(x)|0> \Bigl[{{\partial
\Phi_{\beta}(x)} \over{\partial x_o}}+i\omega_{\bf p_a}
\Phi_{\beta}(x)\Bigr]\eqno(1.6b)$$
 where we have used the following expressions for a one-particle (antiparticle)
 states

$$<0|\Phi_{\beta}(x)|{\bf p}\gamma>=
Z^{-1/2}\int {{d^3p'}\over
{(2\pi)^32\omega_{\bf p'}}}e^{ip'x}<0|a_{{\bf p'}\beta}(x_o)|{\bf p}\gamma>
=Z^{-1/2}\delta_{\gamma\beta}e^{ipx},
\eqno(1.7a)$$

¨

$$<{\bf p}_a\gamma|\Phi_{\beta}(x)|0>=
Z_a^{-1/2}\int {{d^3p'}\over
{(2\pi)^32\omega_{\bf p'}} }e^{ip'x}
<{\bf p}_a\gamma|{b^+}_{{\bf p'}{\gamma}}(x_0)|0>
=Z_a^{-1/2}\delta_{\gamma\beta}e^{-ip_ax},
\eqno(1.7b)$$ here index ${\sl ``a''}$
 denotes the antiparticle state and
  $Z$ is the renormalisation constant.

The field operators $a_{{\bf p}\gamma}(x_0)$ and ${b^+}_{{\bf p}{\gamma}}(x_o)$
are simply defined via the corresponding source operator
$\partial /
\partial{ x_{0}}a_{{\bf p}\beta}(x_0)=i\int d^3x e^{ipx}
j_{\beta}(x)$,
where
$\Bigl({{\partial^2} /{\partial{ x_{\mu}}\partial{x^{\mu} }
}}+m^2\Bigr) \Phi_{\beta}(x)=j_{\beta}(x)$. Moreover,
these operators determine the transition  ${\cal S}$-matrix 

$${\cal S}_{mn}\equiv
<out;{\bf p'}_{1} {\alpha'}_1,...,{\bf p'}_{m}{\alpha'}_m| {\bf
p}_{1} {\alpha}_1,...,{\bf p}_{n}{\alpha}_n;in>=
\prod_{i=1}^m\Bigl[ \int d{{x^0}'}_i {{d}\over{d {x^0}_i}}\Bigr]$$
$$\prod_{j=1}^n\Bigl[ \int d{x^0}_j {{d}\over{d {x^0}_j}}
\Bigr]<0|T\Bigl( a_{{\bf
p'}_m\alpha'_m}({x^0}'_m),...,a_{{\bf p'}_1\alpha'_1}({x^0}'_1)
a_{{\bf p}_n\alpha_n}^+({x^0}_n),...,a_{{\bf
p}_1\alpha_1}^+({x^0}_1)\Bigr)|0>. \eqno(1.8)$$

Next we shall consider the transformations of $\Phi_{\beta}(x)$
 according to the conformal transformations of 
${\Phi_{\gamma}^{(\pm)}}(q)$  

$${\Phi_{\gamma}^{(\pm)}}(q)\to {\Phi_{\gamma}^{(\pm)}}'(q')
={\cal U}(g) {\Phi_{\gamma}^{(\pm)}}(q)  {{\cal U}}^{-1}(g) =
{\cal T}^{\beta}_{\gamma} \Phi_{\beta}^{(\pm)}(g^{-1}q),\eqno(1.9)$$ 
where $g$ indicates one of the (1.1a)-(1.1e) transformations 
$g\equiv\Bigl(T(h),R(\Lambda),{\cal D}(\lambda),{\cal K}(M,h)\Bigr)$,
${\cal T}^{\beta}_{\gamma}$ is the spin-isopin matrix and ${\cal U}(g)$
are defined through the generators of the corresponding transformations
in the well known form:

$$T(h):\ \ \ \ \ {\cal U}(h)=e^{ih_{\mu}{\cal X}^{\mu}};\ \ \
\ \ \biggl[{\cal X}_{\mu},{\Phi_{\gamma}^{(\pm)}} (q)\biggr]
=-i{{\partial} \over{\partial q^{\mu}} }{\Phi_{\gamma}^{(\pm)}}
(q), \eqno(1.10a)$$ ¯

$$R(\Lambda):\ \ \ \ \ {\cal U}(\Lambda)=
e^{i\Lambda_{\mu\nu}{\cal M}^{\mu\nu}};\ \ \ \biggl[{\cal
M}_{\mu\nu},{\Phi_{\gamma}^{(\pm)}} (q)\biggr]
=-i\biggl(q_{\mu}{{\partial} \over{\partial q^{\nu} } }\ - \
q_{\nu}{{\partial} \over{\partial q^{\mu} }
}-i\Sigma_{\mu\nu}\biggr) {\Phi^{(\pm)}}(q) \eqno(1.10b)$$ where
$\Sigma_{\mu\nu}=0$ for scalars,
$\Sigma_{\mu\nu}=i/4[\gamma_{\mu},\gamma_{\nu}]$ for fermions and
$\bigl(\Sigma_{\mu\nu}V_{\rho}\bigr)=ig_{\mu\rho}V_{\nu}-ig_{\nu\rho}V_{\mu}$
for the vectors $V_{\rho}$.

$${\cal D}(\lambda):\ \ \ \ \ {\cal U}(\lambda)=e^{i\lambda D};\ \ \
\ \ \biggl [D,{\Phi_{\gamma}^{(\pm)}} (q)\biggr]
=-i\biggl(q_{\mu}{{\partial} \over{\partial q^{\mu} }
}+id_m\biggr) {\Phi_{\gamma}^{(\pm)}} (q),\eqno(1.10c)$$ 
where $d_m$ indicates  the scale dimension of field. 
For example, in the scale-invariant case $d_m=-3$.
$${\cal K}(M,{\hbar}):\ \ \ \ \ {\cal U}({\hbar})=e^{i{\hbar}_{\mu}K^{\mu}};
\ \ \ \ \ \biggl[ K_{\mu},{\Phi_{\gamma}^{(\pm)}} (q)\biggr]
=-\biggl(2q_{\mu}D-q^2{\cal
X}_{\mu}+2iq^{\nu}\Sigma_{\mu\nu}\biggr) {\Phi_{\gamma}^{(\pm)}}
(q).\eqno(1.10d)$$

According to (1.4b)
the conformal transformations of the operators
 ${\Phi_{\gamma}^{(\pm)}}(q)$ (1.9) generates the corresponding 
transformations of $\Phi_{\gamma}'(x)$

$$\Phi_{\gamma}'(x)={\cal T}^{\beta}_{\gamma}
\int {{d^4 q}\over{(2\pi)^4  } }
\Bigl[{\Phi_{\beta}^{(+)}} (g^{-1}q)e^{-iqx}+
{\Phi_{\beta}^{(-)}}^+ (g^{-1}q)e^{iqx}\Bigr].\eqno(1.11)$$
In particular,  eq.(1.11) consists of the  following transformations:

\par
{\underline {\bf Four-momentum translation:}}

For a charged particle a four-momentum translation
is equivalent to  

$$q'_{\mu}=q_{\mu}+h_{\mu}\ \ \ \ \ \Longrightarrow
i{{\partial}\over{\partial x'_{\mu} }}=
i{{\partial}\over{\partial x_{\mu} }}+h_{\mu},
\eqno(1.12a)$$
which implies the well known gauge transformation of the charged particle
field operator  
$${\Phi_{\gamma}}'(x)=e^{ihx}\Phi_{\gamma}(x).\eqno(1.12b)$$

In order to get the  the gauge transformation 
formula (1.12b)  we introduce
 the following transformations
of ${\Phi_{\gamma}^{(\pm)}}(q)$

$${\Phi_{\gamma}^{(+)}}' (q)={\Phi_{\gamma}^{(+)}}(q+h);\ \ \
{{\Phi_{\gamma}^{(+)}}^+}' (q)={\Phi_{\gamma}^{(+)}}^+(q+h)\eqno(1.13a)$$
¨
$$ {\Phi_{\gamma}^{(-)}}' (q)={\Phi_{\gamma}^{(-)}}(q-h);\ \ \
{{\Phi_{\gamma}^{(-)}}^+}' (q)={\Phi_{\gamma}^{(-)}}^+(q-h).\eqno(1.13b)$$
After substitution of (1.13a,b) in (1.4b) we get

$${\Phi_{\gamma}}'(x)=\int {{d^4 q}\over{(2\pi)^4 } }
\Bigl[ {\Phi_{\gamma}^{(+)}}(q+h)e^{-iqx}+{\Phi_{\gamma}^{(-)}}^+(q-h)
e^{iqx}\Bigr]=e^{ihx}\Phi_{\gamma}(x)
\eqno(1.14).$$
In the same way we obtain

$$\int {{dq_o}\over{2\pi}} {\Phi_{\gamma}^{(+)}} (q_o+h_o,{\bf p+h}) 
e^{-iq_0x_0}=
\int {{dq_o}\over{2\pi}} {\Phi_{\gamma}^{(+)}} (q_o,{\bf p+h}) e^{-i(q_0-h_o)
x_0}$$

$$={{e^{-i\omega_{\bf p+h}x_o}}\over{2\omega_{\bf p+h}}}
a_{\bf p+h\gamma}(x_o) e^{ih_ox_o}
=Z^{1/2}
{{e^{i{\bf p+h}x}}\over{2\omega_{\bf p+h}}}
\sum_{\beta}
<0|\Phi'_{\gamma}(x)|{\bf p+h}\beta>
a_{\bf p+h\beta}'(x_o),
\eqno(1.15a)$$
where the operator
$$a_{\bf p\gamma}'(x_o)=i
 \sum_{\beta}\int d^3x <0|\Phi'_{\beta}(x)|{\bf p}\gamma>
{{ \stackrel{\longleftrightarrow}{\partial}}\over{\partial x^o}}
 \Phi'_{\beta}(x)\eqno(1.15b)$$
coincides with the operator (1.5b)
$$a_{\bf p\gamma}'(x_o)=a_{\bf p\gamma}(x_o).\eqno(1.15c)$$
Thus the gauge transformation (1.12a,b) generates
the following transformation of
the ${\cal S}$-matrix (1.8) 

$${\cal S'}_{mn}=
<out;{\bf p'_1+h}\ {\alpha'}_1,...,{\bf p'_m+h}\ {\alpha'}_m| {\bf
p_1+h}\ {\alpha}_1,...,{\bf p_n+h}\ {\alpha}_n;in>
\eqno(1.16)$$
This expression differs from ${\cal S}_{mn}$ by a shift of the position 
of the origin in the 3D momentum space. 
Therefore ${\cal S'}_{mn}={\cal S}_{mn}$
because only the relative momenta are physically
meaningful. 
A more complicated shift of a four-momentum operator 
${\widehat P}_{\mu}'={\widehat P}_{\mu}-<0|{\widehat P}_{\mu}|0>$
is often used in quantum field theory concerning the so-called 
zero-mode problem (see for example ch. 12 of \cite{BD}).

{\underline {\bf For\ a\ neutral\ particle }} field operator
$\phi_{\gamma}(x)$ the
 translation  $q_{\mu}'=q_{\mu}+h_{\mu}$ (1.12a) has a more complicated 
form due 
to absence of the antiparticle degree of freedom. In particular,
using (1.13a) we obtain 

$$\phi_{\gamma}'(x)=\int {{d^4 q}\over{(2\pi)^4 } } \Bigl[
{\phi_{\gamma}^{(+)}} (q)e^{-i(q-h)x}+ {\phi_{\gamma}^{(+)}}^+(q)
e^{i(q-h)x}\Bigr]\ne e^{ihx}\phi_{\gamma}(x)\eqno(1.17a)$$
or
$$\phi_{\gamma}'(x)=\int {{d^4 q}\over{(2\pi)^3}}
\delta((q_o-h_o)^2-({\bf q-h})^2-m^2)\theta(q_o-h_o)$$
$$\Bigl[
{\sl a}'_{{\bf q}{\gamma}}(x_0)e^{-i(\omega_{\bf q-h}-h_o)x_o+i({\bf q-h})
{\bf x} }+{{{\sl a}'}^+}_{{\bf q}{\gamma}}(x_o)
e^{i(\omega_{\bf q-h}-h_o)x_o+i({\bf q-h}){\bf x} }\Bigr],
\eqno(1.17b)$$

where

$${\sl a}_{\bf p\gamma}'(x_o)=i
 \sum_{\beta}\int d^3x <0|\phi'_{\beta}(x)|{\bf p}\gamma>
{{ \stackrel{\longleftrightarrow}{\partial}}\over{\partial x^0}}
 \phi'_{\beta}(x)\eqno(1.18)$$

According to (1.17 a,b) $\phi_{\gamma}'(x)$ remains to be  hermitian
after translations
$q_{\mu}'=q_{\mu}+h_{\mu}$. On the other hand these
translations generate the nontrivial dependence of
$\phi_{\gamma}'(x)$ on  $h_{\mu}$.
The similar dependence on the additional parameter $h_{\mu}$ 
appears in the real fields  $\phi_{1,2}(x)$
  constructed from the charged pion fields
 $\pi_{\pm}(x)$ after their gauge transformation (1.12a,b)
${\phi'}_{1}(x)=1/\sqrt{2}\Bigl(
\exp{(-ihx)}\pi_{+}(x)+\exp{(ihx)}{\pi_{+}}^+(x)
\Bigr)$ and

${\phi'}_{2}(x)=i/\sqrt{2}\Bigl(
\exp{(-ihx)}\pi_{+}(x)-\exp{(ihx)}{\pi_{+}}^+(x)
\Bigr)$.
It must be noted, that
a splitting of $\phi_{\gamma}(x)$ on the positive and the negative
frequency parts 
$\phi_{\gamma}(x)=\phi_{\gamma}^{(+)}(x)+{\phi_{\gamma}^{(+)}}^{\dagger }(x)$
can be realized 
with arbitrary
parameter $\alpha$ \cite{Weinberg}
as $\phi_{\gamma}(x)= e^{i\alpha}\phi_{\gamma}^{(+)}(x)+
e^{-i\alpha}{\phi_{\gamma}^{(+)}}^{\dagger }(x)$. 
In our case the additional dependence
of $\phi_{\gamma}´(x)$ on $h_{\mu}$ is result of the
condition (1.13a) which is necessary for the gauge
transformations rule (1.12a,b) of the charged field operators.

Using the normalization condition for functions
$f_{p-h}(x)=e^{i(p_o-h_o)x_o-i({\bf p-h}){\bf x} }$ 

$$i\int f_{p'-h}^{\ast}(x)
{{\stackrel{\longleftrightarrow}{\partial} }\over{\partial x^0}}
f_{p-h}(x)d^3x=2(p_o-h_o)
(2\pi)^3\delta({\bf p'-p}),\eqno(1.19a)$$

$$
i\int f_{p'-h}(x)
{{\stackrel{\longleftrightarrow}{\partial}   }\over{\partial x^o}}
f_{p-h}(x)d^3x=i\int f_{p'-h}^{\ast}(x)
{{ \stackrel{\longleftrightarrow}{\partial}}\over{\partial x^0}}
f_{p-h}^{\ast}(x)d^3x=0\eqno(1.19b)$$

It is easy to obtain
$${\sl a}_{\bf p\gamma}'(x_o)=a_{\bf p\gamma}(x_o), \eqno(1.20)$$
 where $a_{\bf p\gamma}(x_o)=i \sum_{\beta}\int d^3x
<0|\phi_{\beta}(x)|{\bf p}\gamma> {{
\stackrel{\longleftrightarrow}{\partial}}/{\partial x^0}}
 \phi_{\beta}(x)$. Relation (1.20) is similar to the relation (1.15c)
for the charged fields.
This means, that ${\cal S}$-matrix 
transforms according to the same relation (1.16)
for the charged and neutral particles
 after translation in the 4D momentum space 
$q'_{\mu}=q_{\mu}+h_{\mu}$.
The dependence on the dummy variables $q_{o}$ and $q_{o}+h_{o}$ 
 disappears 
in the ${\cal S}$-matrix 
after the appropriate integration in (1.5a), (1.6a)
and (1.15a,b). 
 Thus for the ${\cal S}$-matrix and other observables the translation of 
$q\equiv(q_o,{\bf p})$ is reduced to the 3D translations
   ${\bf p'=p+h}$ which does not affect these observables.

{\underline{\bf Rotation (1.1b) and dilatation (1.1c) of $q_{\mu}$}}  
for the particle field operator $\Phi_{\gamma}(x)$ (1.11)
may be performed
using the rotations (1.10b)  and scale transformations
 (1.10c) of  ${\Phi_{\gamma}^{(\pm)}} (q)$ operators.
In particular, rotations $q'_{\mu}=\Lambda_{\mu\nu}q^{\nu}$
generates the following transformation of the field operators in the
configuration space
$$R(\Lambda):\ \ \ \ \
{\Phi_{\gamma}'}(x_{\mu})=\Phi_{\gamma}
(\Lambda_{\mu\nu}^{-1}x^{\nu}),\eqno(1.21)$$
and for dilatation $q_{\mu}'=e^{\lambda} \ q_{\mu}$   we have
$${\cal D}(\lambda):\ \ \ \ \
{\Phi_{\gamma}'}(x)=e^{4\lambda}\Phi_{\gamma}(e^{-\lambda}\ x).
\eqno(1.22)$$
Therefore the rotations and dilatation of $\Phi_{\gamma}(q)$
generate the analogical transformations of $\Phi_{\gamma}(x)$.

{\underline{\bf Special conformal transformation and inversion:}}
Special conformal transformation of  $q_{\mu}$ (1.1e) for  
 ${\Phi_{\gamma}^{(\pm)}}(q)$ has the form

$${\Phi_{\gamma}^{(+)}}'(q)={\Phi_{\gamma}^{(+)}}\Bigl((q^I+h)^I\Bigr);
\ \ \
{{\Phi_{\gamma}^{(+)}}^+}'(q)={\Phi_{\gamma}^{(+)}}^+\Bigl((q^I+h)^I\Bigr)
\eqno(1.23a)$$
¨
$$ {\Phi_{\gamma}^{(-)}}' (q)={\Phi_{\gamma}^{(-)}}\Bigl((q^I-h)^I\Bigr);\ \ \
{{\Phi_{\gamma}^{(-)}}^+}' (q)={\Phi_{\gamma}^{(-)}}^+\Bigl((q^I-h)^I\Bigr),
\eqno(1.23b)$$
where the index $^I$ relates to the inversion of $q_{\mu}$. 
According to (1.11) we get

$$\Phi_{\gamma}'(x)=
\int {{d^4 q}\over{(2\pi)^4 } }
\Bigl[{\Phi_{\gamma}^{(+)}}\Bigl((q^I+h)^I\Bigr)e^{-iqx}+
{\Phi_{\gamma}^{(-)}}^+\Bigl((q^I-h)^I\Bigr)e^{iqx}\Bigr].
\eqno(1.24)$$

Arbitrary operator ${\Phi_{\gamma}^{(\pm)}}(q)$ may be divided into 
inversion invariant and inversion anti-invariant parts

$${\Phi_{\gamma}^{(\pm)}}_{inv.}(q)={1\over 2}
\Bigl[{\Phi_{\gamma}^{(\pm)}}(q)+
{\Phi_{\gamma}^{(\pm)}}(q^I) \Bigr]\eqno(1.25a)$$

and

$${\Phi_{\gamma}^{(\pm)}}_{anti-inv.}(q)={1\over 2}
\Bigl[{\Phi_{\gamma}^{(\pm)}}(q)-
{\Phi_{\gamma}^{(\pm)}}(q^I) \Bigr].\eqno(1.25b)$$

Expressions (1.25a,b) simplifies
use of the special conformal transformation.

\vspace{0.7cm}

\centerline{\bf{2.\  Five dimensional projection}}

\vspace{0.4cm}

\par
The invariant form of the $O(2,4)$ group 
$\kappa_A\kappa^A=0$ (1.2) can be represented in the five dimensional form
with  $q_{\mu}$ (1.3) variables

$$q_{\mu}q^{\mu}+M^2{{\kappa_{-}}\over{\kappa_{+}}}
=0,\eqno(2.1a)$$
where
$$q_{\mu}={{\kappa_{\mu}}\over{\kappa_+}};\ \ \ \
\kappa_{\pm}={{ \kappa_5\pm\kappa_6}\over {M}}. \eqno(2.1b)$$
It is convenient to introduce new fifth momentum $q_5$
instead of two variables
 $\kappa_{\pm}$ (or
$\kappa_5$, $\kappa_6$) in (2.1a).
This procedure implies  a projection
of the 6D rotational invariant form
$\kappa_A\kappa^A=0$  into corresponding 5D forms.  
There  exists only two
 5D  De Sitter spaces with the constant curvature which
have the  invariant forms of the 
$O(2,3)$ and $O(1,4)$ rotational groups \cite{BK,Todorov,K2} 

$$q_{\mu}q^{\mu}+q_5^2=M^2 \ \ \ \ \ \ \ \ \ \ \ \ \ \ \ \ \ \ \ \ \ \
 \ \ q_5^2 =M^2 { {2\kappa_{5}} \over {\kappa_{5}+\kappa_{6} } },\eqno(2.2a)$$

and

$$q_{\mu}q^{\mu}-q_5^2=-M^2 \ \ \ \ \ \ \ \ \ \ \ \ \ \ \ \ \ \ \ \ \ \ \
q_5^2 =M^2 { {2\kappa_{6}}\over {\kappa_{5}+\kappa_{6} } }.\eqno(2.2b)$$

\par
\footnotetext{ In the literature often is considered the stereographic  
projection of the 6D cone $\xi_A\xi^A=0$  into 4D
Minkowski space  with coordinates
$x_{\mu}=\xi_{\mu}{\ell}/(\xi_5+\xi_6)$, where
at the intermediate stage  are used
projections on the 5D hyperboloid
 $\eta_{\mu}\eta^{\mu}-\eta_5^2=- {\ell}^2$ (see for example
ch. 13 of \cite{IZ})  with  
 $\eta_{\mu}=\xi_{\mu}{\ell}/\xi_5$; $\eta_{5}=\xi_{6}{\ell}/\xi_5$ 
and
 $\eta_{\mu}=2x_{\mu}/(1-x^2/{\ell}^2);\ \ \ \eta_{5}={\ell}
 (1+x^2/{\ell}^2)/(1-x^2/{\ell}^2)$.
Here
$x^2=\ell^2(\eta_5/\ell-1)/(\eta_5/\ell+1)$ and at first sight
$x_{\mu}$ is  not restricted
by the 5D condition  $\eta_{\mu}\eta^{\mu}-\eta_5^2=- {\ell}^2$
like $q^2$  (2.2a,b)  in table  1 or 2.
Nevertheless, the 6D invariant form can be rewritten as
$x^{\mu}x_{\mu}+\ell^2(\xi_5-\xi_6)/(\xi_5+\xi_6)=0$ and the
appropriate projection  into  5D hyperboloid
$x^{\mu}x_{\mu}\pm x_5^2=\pm\ell^2$
with $x_5^2=2\xi_5(\ or\ \xi_6)\ell^2/(\xi_5+\xi_6)$
generates the corresponding restrictions.}

The real variables $q_{\mu}$ and $q_5$    
 are defined in the regions
$(-\infty,+\infty)$ and  $[0,+\infty)$ respecti\-vely
\footnotemark.
In the considered formulation
 $q_{\mu}$ and $q_5$ are disposed on  hyperboloids
(2.2a,b). In particular, we can place $0<q^2\le M^2$ 
 and $q^2>  M^2$
on the hyperboloids  (2.2a) and (2.2b) respectively  Thus 
the conformal transformations for the whole $q^2$ values may be performed
 using  both hyperboloid  (2.2a,b). 
The values of $q_{\mu}$ and $q_5$ on these hyperboloids 
are singlevalued connected with each other
via inversion 
$q_{\mu}'=-M^2 q_{\mu}/q^2$ (1.1d). On the 6D cone 
$\kappa_A\kappa^A=0$ (1.2) inversion (1.1d) 
 can be carried out using the reflection of the    
$\kappa_6$ variable

$$I(M^2):\ \ \ \ \ \kappa_5^I=\kappa_5,\ \ \ \kappa_6^I=-\kappa_6;\ \ \
 ¨«¨\ \ \
\kappa_+^I=\kappa_-,\ \ \ \kappa_-^I=\kappa_+,\eqno(2.3)$$ 
which generates $q_{\mu}^I=-M^2 q_{\mu}/q^2$ according to
(2.1a,b). The advantage of  the 6D representation (2.3)
of the 4D transformation  $q_{\mu}^I=-M^2 q_{\mu}/q^2$ is that it
determines the transparent realization of the nonlinear 4D
transformation using the simple reflection in 6D space. 
In particular, for $0<q_{\mu}\le M^2$ on the hyperboloid
  $q^2+q_5^2=M^2$ , we have
${q^2}^I=M^4/q^2\ge M^2$ and
 $(-{q_5^2}+M^2)^I=
M^2 {\kappa_-}^I/{\kappa_+}^I =M^2/(\kappa_-/\kappa_+)
=M^4/({q_5}^2-M^2)$. Therefore, if $q_{\mu}^I$ belongs to  
 hyperboloid ${q^2}^I+(-q_5^2+M^2)^I=0$ (2.2b), then
$q_{\mu}$ will be placed on the other hyperboloid, because
${q^2}^I+(-q_5^2+M^2)^I=
M^4/[q^2(q_5^2-M^2)](q^2+q_5^2-M^2)=0$. 
The distribution of the  regions of the 
5D hyperboloid $q^2 \pm q_5^2=\pm M^2$
(2.2a,b) which covers the whole values of $q_{\mu}$ and $q_5$
 ($-\infty < q^2 < \infty$ and $q_5^2\ge 0$) 
 is given in table 1.

\par
\footnotetext{ The border point $q^2=0$ with
$q_5^2=M^2$ is included in the
domain $q_{\mu}q^{\mu}+
q_5^2=M^2$, because it belongs to the physical spectrum of the
massless particles.
After inversion $q^2=0$ 
transforms into $q^2=\infty$ of the hyperboloid  $q_{\mu}q^{\mu} -
q_5^2=-M^2$  in the region $II$.}

\vspace{0.7cm}

\centerline{\bf Table  1}
\vspace{0.2cm}

\hspace{-0.75cm}
\begin{tabular}{|c|c|c|c|c|} \hline\hline
        &        {\bf I}            &         {\bf II}
            &      {\bf III}            &         {\bf IV }
\\    \hline
           & $q^2+q_5^2=M^2$ & $q^2-q_5^2=-M^2$ & $q^2+q_5^2=M^2$
& $q^2-q_5^2=-M^2$ \\ \hline
 $ q^2$     &   $0\le q^2 \le M^2$   & $M^2< q^2< \infty$ &$-\infty<q^2<-M^2$
& $-M^2\le q^2< 0 $\\  \hline
 $ q_5^2$       &   $0\le q_5^2 \le M^2$   & $2M^2\le q_5^2< \infty$ 
                & $2M^2< q_5^2< \infty$ &
 $0\le q_5^2< M^2$ \\ \hline \hline
\end{tabular}
\vspace{0.5cm}

\par
 
Momentum   $q_{\mu}$ from the region $I$
 is singlevalued connected 
with  $q_{\mu}$  in the region   $II$ 
  via inversion 
$\Bigl\{ {q_{\mu}} \Bigr\}_{II\ region}
=-M^2\Bigl\{q_{\mu}/q^2 \Bigr\}_{I\ region}$ 
and vice versa
$\Bigl\{ {q_{\mu}} \Bigr\}_{I\ region}
=-M^2\Bigl\{q_{\mu}/q^2 \Bigr\}_{II\ region}$.
In the same manner are connected the four momenta in the regions 
$III$ and $IV$ with  $q^2<0$.
For $M\to\infty$ 5D spaces transforms into ordinary Minkowski space
with the domains $I$ and $IV$.
For  $M\to 0$  we get again a Minkowski space
with remained regions $II$ and $III$.

\par
The scale transformation have the different form
 in the different areas in table 1.
In 6D space the scale transformation $q_{\mu}'=e^{\lambda}q_{\mu}$ (1.1c)
 implies the rotation in the
 (6,5) plane. For $q^2\ge 0$ rotation
 $\kappa_5=M\ sh(\lambda);\kappa_6=M\ ch(\lambda)$
generates the following transformation of
$q^2$:\ \ \ $  q_{\mu}q^{\mu}=-
M^2{({\kappa_5-\kappa_6})/({\kappa_5+\kappa_6})
}=M^2e^{-2\lambda}$. For negative $q^2<0$  
we take $\kappa_5= M\ ch(\lambda);\
\kappa_6=M\ sh(\lambda)$  (i.e. $\kappa_+=e^{\lambda}$)
which gives
$q_{\mu}q^{\mu}=-M^2e^{2\lambda}$. 
The corresponding transformation of
$q_5^2$ with the related  $\lambda$, is given in table 2.


\newpage
\centerline{\bf Table 2}
\vspace{0.2cm}

\hspace{-1.5cm}
\begin{tabular}{|c|c|c|c|c|} \hline\hline
 {\rm rotation}          & \multicolumn{2}{|c}
{$\kappa_5=M\ sh\lambda;\  \kappa_6=M\ ch\lambda$}
& \multicolumn{2}{|c|}{$\kappa_5=M\ ch\lambda; \ \kappa_6=M\ sh\lambda$} \\
\hline
$\lambda$           &  $\lambda>0$ &  $\lambda<0$ & $\lambda<0$ &
$\lambda>0$  \\ \hline
        &        {\bf I}            &         {\bf II}
            &      {\bf III}            &         {\bf IV }
\\    \hline
{\rm hyperboloid} & $q^2+q_5^2=M^2$ & $q^2-q_5^2=-M^2$ & $q^2+q_5^2=M^2$
& $q^2-q_5^2=-M^2$ \\ \hline
 $ q^2$     &   $q^2=M^2e^{-2\lambda}$   & $q^2=M^2e^{-2\lambda}$ &
$q^2=-M^2e^{-2\lambda}$
& $q^2=-M^2e^{-2\lambda}$\\  \hline
 $ q_5^2$       &   $q_5^2=M^2(1-e^{-2\lambda})$   &
$q_5^2=M^2(1+e^{-2\lambda})$ & $q_5^2=M^2(1+e^{-2\lambda})$ &
 $q_5^2=M^2(1-e^{-2\lambda})$ \\ \hline \hline
\end{tabular}
\vspace{0.5cm}

\par
In table 2 it is shown, that the scale transformation 
parameter $\lambda$ (or $\kappa_+=e^{-\lambda}$) single-valued determines
$q_5^2$ and $q^2$.
In particular, in the region I
with $q^2+q_5^2=M^2$
the scale transformation  is realizable by $\lambda>0$,
which implies the compression of $q^2$.
In opposite to this,  in the region II with $q^2-q_5^2=-M^2$ we have 
  $\lambda<0$, i.e.
the same scale transformation generates the stretching
of $q^2$.  
A similar scale transformation can be observed for $q^2<0$, where
in the region III dilatation generates stretching and in the region IV we
have compression. 
In other words, 
projections of the 5D cone  $\kappa_A \kappa^A=0$
on the 5D hyperboloid for $q^2>0$ implies
$$\kappa_A \kappa^A=0\Longrightarrow
q_{\mu}q^{\mu}+q_5^2=M^2,\ \ \ \  for  \ \ 0\le q_{\mu}q^{\mu}\le M^2\ \
with\ \ \ \lambda\geq 0\eqno(2.4a)$$
$$\kappa_A \kappa^A=0\Longrightarrow
q_{\mu}q^{\mu}-q_5^2=-M^2,\ \ \ \  for  \ \ q_{\mu}q^{\mu}> M^2\ \ 
with\ \ \ \lambda\le 0.\eqno(2.4b)$$

Inversion ${q'}^2=M^4/q^2$  replaces
the points from the internal region $0\le q^2<M^2$ (section I)
with the points from the external region $q^2>M^2$ (section II)
and vice versa. Thereby  
the hyperboloid  $q^2+q_5^2=M^2$  we shall  denote 
as the ``internal'' and the
hyperboloid  $q^2+q_5^2=M^2$  we shall call as the 
``external''.

\par
By translation $q_{\mu}'=q_{\mu}+h_{\mu}$ the 6D cone
 $\kappa_A\kappa^A=0$, as well as the 5D forms
(2.2a,b) $q^2\pm q_5^2=\pm M^2$ are preserved.
In particular, 
after the appropriate 6D rotations 
$\kappa_{\mu}'=\kappa_{\mu}+h_{\mu}\kappa_{+};$
$\kappa_{+}'=\kappa_{+};$ $\kappa_{-}'=\kappa_{-}+2/M^2
h_{\nu}\kappa^{\nu}+h^2/M^2\kappa_+$, we get
$${q^2}'=q^2+2 h_{\nu}q^{\nu}+h^2=-M^2{{\kappa_{-}'}\over{\kappa_{+}}};\ \ \
 \ \ {q_5^2}'=q_5^2\pm(2 h_{\nu}q^{\nu}+h^2),\eqno(2.5)$$
where the sign $-$ corresponds to
  $q^2+q_5^2=M^2$ and $+$ relates
to  $q^2-q_5^2=-M^2$. Using (2.5) we have 
${q^2}'\pm{q_5^2}'={q^2}\pm{q_5^2}$. 
Nevertheless, the transformation
$q_{\mu}'=q_{\mu}+h_{\mu}$ can generate a transition from the
time-like region
$q^o\ge 0$ into space-like region $q^o<0$.
Transition between the $q^2> 0$ and $q^2< 0$ regions
can be compensated by inversion or by transposition of  $\kappa^{6}$ and
$\kappa^{5}$ variables.

\par
It must be noted, that inversion transforms the generators of the
conformal group  (1.10a) - (1.10d)
in the following way $\  ^{\dagger}$\footnotemark
$${\cal X}_{\mu}=I(M^2)\ K_{\mu}\ I(M^2);\ \ \
{\cal M}_{\mu\nu}=I(M^2)\ {\cal M}_{\mu\nu}\ I(M^2);$$
$$\ \ \ D'=I(M^2)\ D\ I(M^2)=-D;\ \ \ K_{\mu}=I(M^2)\ {\cal X}_{\mu}\ I(M^2),
\eqno(2.6)$$

Therefore one can perform the conformal transformations 
only in the ``internal'' regions $I$ and $III$ and
 obtain the corresponding transformations in
the ``external'' regions $II$ and $IV$ using the inversion.

\par
\footnotetext{
A similar transformation  can be performed using
the Weyl reflection, i.e. rotation through $90^o$ in the $(0,5)$
plane \cite{BR}.}

\vspace{0.5cm}

{\underline {\bf 5D reduction of the field operators:}}\ \ \ \ \
Next we have to connect a 6D field operator
$\phi^{(\pm)}(\kappa_A)$ d
 ($A=\mu;5,6\equiv$ 0, 1, 2, 3; 5,6)
with the 5D operators
${\phi^{(\pm)}_{inr}}(q,q_5)$  and
${\phi^{(\pm)}_{ext}}(q,q_5)$.
Here index $\  ^{(\pm)}$ corresponds to positive or negative frequency,
the subscripts  $inr$ or
$ext$ indicate the surfaces   $q^2+q_5^2=M^2$ and $q^2+q_5^2=-M^2$
correspondingly. Afterwards for the sake of simplicity we  omit the
spin-isospin indices ${\gamma}$.

\par
 According to the manifestly covariant construction of the  $O(2,4)$ 
conformal group \cite{M1,M2}, we shall use following independent variables

$$q_{\mu}=\kappa_{\mu}/\kappa_{+};\ \ \ \kappa_{+}=(\kappa_{5}+\kappa_{6})/M;
\ \ \ \kappa^2=\kappa^A \kappa_A.\eqno(2.7)$$
Only this choice of the  variables
makes  independent the generators of the conformal group $O(2,4)$
on  $\partial/\partial{\kappa^2}$.
In particular, for a spinless particle these generators are
$$ {\cal X}_{\mu}=i{{\partial}\over{\partial q^{\mu}}};\ \ \
{\cal M}_{\mu\nu}=i\Bigl(q_{\mu}{{\partial}\over{\partial q^{\nu}}}-
q_{\nu}{{\partial}\over{\partial q^{\mu}}}\Bigr);$$

$$D=i\Bigl(q_{\mu}{{\partial}\over{\partial
q^{\mu}}}-k_{+}{{\partial}\over{\partial k_{+}}}\Bigr);\ \ \
{\cal K}_{\mu}=2q_{\mu}D-q^2{\cal X}_{\mu}.\eqno(2.8)$$

Using the variables (2.7) the 6D field operator takes the form
$$\phi^{(\pm)}(\kappa_A)\equiv\phi^{(\pm)}(q_{\mu},\kappa_+,\kappa^2).
\eqno(2.9)$$
The homogeneous over the scale variable $\kappa_+$ operator
$\phi^{(\pm)}(q_{\mu},\kappa_+,\kappa^2)$  may be rewritten as
$$\phi^{(\pm)}(q_{\mu},\kappa_+,\kappa^2)= (\kappa_+)^d
\varphi^{(\pm)}(q_{\mu},\kappa^2),\eqno(2.10)$$
and for the 4D physical field operator. in analogue 
to \cite{M1,M2} we get
$$\Phi^{(\pm)}(q_{\mu})= (\kappa_+)^{-d}{\cal O}
\phi^{(\pm)}(\kappa_A),\eqno(2.11)$$
where $d$ defines the scale dimension of the considered operator,
and ${\cal O}$ 
acts on the spin-isospin variables.

\par
In the present paper we shall use other  recipe
of projection of the 6D cone $\kappa^2=0$ 
 into 5D surfaces
$q_{\mu}q^{\mu}\pm q_5^2=\pm M^2$ and in the 4D momentum space.
In particular, we shall treat the condition  $\kappa^2=0$ as the
dynamical restriction, i. e. we shall require the validity of the
following constraint
$$\Bigl(\kappa^A\kappa_A\Bigr)\phi^{(\pm)}(q_{\mu},\kappa_+,\kappa^2)=
\kappa_{+}^2\Bigl(q_{\mu}q^{\mu}+M^2{{\kappa_{-}}\over{\kappa_{+}}}\Bigr)
\phi^{(\pm)}(q_{\mu},\kappa_+,\kappa^2)=0,
\eqno(2.12)$$
Projection of this equation on the 5D surfaces
 $q_{\mu}q^{\mu}\pm q^2_5=\pm M^2$ gives

$$\Bigl(q_{\mu}q^{\mu}+q_5^2-M^2\Bigr)\phi^{(\pm)}_{inr}
\Bigl(q_{\mu},\kappa_+,{\kappa_+}^2(q_{\mu}q^{\mu}+q_5^2-M^2)
\Bigr)=0,\eqno(2.13a)$$

$$\Bigl(q_{\mu}q^{\mu}-q_5^2+M^2\Bigr)\phi^{(\pm)}_{ext}
\Bigl(q_{\mu},\kappa_+,{\kappa_+}^2(q_{\mu}q^{\mu}-q_5^2+M^2)
\Bigr)=0.\eqno(2.13b)$$

As it was show in table 2, the magnitude of the scale parameter
$\kappa_+= e^{\lambda}$ is unambiguously defined via
$q^2$ or $q_5^2$ variables. Therefore we can introduce the 
5D fields
$$\varphi^{(\pm)}_{inr}(q_{\mu},q_5)\equiv (\kappa_+)^{-d}{\cal O}
\phi^{(\pm)}_{inr}\Bigl(q_{\mu},\kappa_+,{\kappa_+}^2
(q_{\mu}q^{\mu}+q_5^2-M^2)\Bigr),\eqno(2.14a)$$

$$\varphi^{(\pm)}_{ext}(q_{\mu},q_5)\equiv (\kappa_+)^{-d}{\cal O}
\phi^{(\pm)}_{ext}\Bigl(q_{\mu},\kappa_+,{\kappa_+}^2
(q_{\mu}q^{\mu}-q_5^2+M^2)\Bigr).\eqno(2.14b)$$

Then using  eq.(2.13a,b) we get 

$$\Bigl(q_{\mu}q^{\mu}+q_5^2-M^2\Bigr)\varphi^{(\pm)}_{inr}(q_{\mu},q_5)=0;
\eqno(2.15a)$$

$$\Bigl(q_{\mu}q^{\mu}-q_5^2+M^2\Bigr)\varphi^{(\pm)}_{ext}(q_{\mu},q_5)=0;
.\eqno(2.15b)$$

Equations (2.15a,b) present the desired 5D projections of the 6D
constraint (2.12). These relations can be treated also as the 4D equations,
because the fifth momentum on 
 $q^2 \pm q_5^2=\pm M^2$ shell is given 
$q_5=\pm \sqrt{|M^2 \mp q^2|}$.

\vspace{0.5cm}
\begin{center}

{\bf{3.\ 4D and 5D  equations of motion.}}
\end{center}
\vspace{0.4cm}
\par
It is well known that the systems with any dimensional 
parameters are conformal non-invariant. 
Nevertheless one can always perform the conformal transformations 
for the conformal-non-invariant systems.
Conformal transformations  in momentum space can be realized as 
6D rotations  with invariant form  $\kappa^A\kappa_A$
which generates the requirement $\kappa^A\kappa_A\phi(\kappa)=0$ (2.12). 
We shall consider this condition
and the appropriate 5D projections 
$\Bigl(q_{\mu}q^{\mu}\pm{q_5}^2\mp M^2\Bigr)
\varphi^{(\pm)}_{inr,ext}(q_{\mu},q_5)=0$ (2.15a,b)
as restrictions which can be  taken into account 
by construction of the 4D equation of motion 

$$\biggl({{\partial^2}\over{\partial x^{\mu}\partial x_{\mu}}}+
m^2\biggr)\Phi(x)=J(x),\eqno(3.1a)$$
where
$$ J(x)=1/(2\pi)^{4}\int d^4q
\Bigl[e^{-iqx}J^{(+)}(q)+e^{iqx}J^{(-)}(q)\Bigr],\eqno(3.1b)$$ 
or
$$J(x)=\int {{d^3 p}\over{(2\pi)^3 2\omega_{{\bf p}}} }
\Bigl[e^{-ipx}{{\partial}\over{\partial x_0}} a_{{\bf
p}{\gamma}}(x_0)+e^{ipx}{{\partial}\over{\partial
x_0}}{b^+}_{{\bf p}{\gamma}}(x_0) \Bigr]; \ \ \ p_o\equiv
\omega_{{\bf p}} =\sqrt{ {\bf p}^2+m^2}.\eqno(3.1c)$$

\par
In order to determine the relation between 4D and 5D formulations 
we introduce the following boundary conditions over the  fifth
coordinate $x_5$
$$\Phi(x)=\Phi(x,x_5=t_5)=\varphi_{inr}(x,x_5=t_5)+\varphi_{ext}(x,x_5=t_5)
\eqno(3.2a)$$ ¨
$${i\over M}{ {\partial}\over { \partial x_5}}\Phi(x,x_5)|_{x_5=t_5}
=\sum_{a=1,2}{i\over M}{ {\partial}\over { \partial
x_5}}\varphi_{a}(x,x_5) |_{x_5=t_5},\ \ \ where \ \ a=1,2\equiv inr,ext 
\eqno(3.2b)$$
and  for $t_5=x_5$ it is convenient
to choose boundary condition as 
$t_5=\tau=\sqrt{x_o^2-{\bf x}^2}$ or $t_5=0$.

Using (2.15a,b) we get 
$$\biggl({{\partial^2}\over{\partial x^{\mu}\partial x_{\mu}}}+
{{\partial^2}\over{\partial x^5\partial
x_5}}+M^2\biggr)\varphi_{inr}(x,x_5)=0 \eqno(3.3a)$$
 for the internal  hyperboloid with
 the regions $I,III$ in table 1 and ¨
$$\biggl({{\partial^2}\over{\partial x^{\mu}\partial x_{\mu}}}-
{{\partial^2}\over{\partial x^5\partial
x_5}}-M^2\biggr)\varphi_{ext}(x,x_5)=0 \eqno(3.3b)$$
for the ``external'' regions  $II,IV$ \footnotemark.

\footnotetext{
The similar boundary conditions can be introduced 
using the conformal transformation group in the coordinate space.
In particular, 
the 6D invariant form $\xi_A\xi^A=0$ with
$x_{\mu}=\xi_{\mu}{\ell}/(\xi_5+\xi_6)$ and
$x^{\mu}x_{\mu}=-\ell^2(\xi_5-\xi_6)/(\xi_5+\xi_6)$ can be
projected into two 5D hyperboloid
$x^{\mu}x_{\mu}\pm x_5^2=\pm\ell^2$
with $x_5^2=2\xi_5(\ or\ \xi_6)\ell^2/(\xi_5+\xi_6)$.
Then we get the internal  and the external 5D regions with the boundary values 
at $x^2=0,\pm \ell^2$ as it is given for 
$q^2,q_5^2$ variables in table 1.
For $\Phi(x)$ we can introduce an analogue to (3.2a,b)
boundary conditions
$\Phi(x)=\varphi_{inr}(x,x_5=t)+\varphi_{ext}(x,x_5=t)$.
In the such constructions 
the  operator \cite{Gatto} ${ \cal M}_{\mu\nu}(x)
=g_{\mu\nu}-2x_{\mu}x_{\nu}/x^2={\cal
M}_{\mu\nu}(1/x)$  has the properties of the metric tensor
${\cal M}_{\mu\nu}(x){\cal M}^{\nu\sigma}(x)=\delta_{\mu}^{\sigma}$, ${\cal
M}_{\mu\nu}(x)x^{\nu}=-x_{\mu}$ and $\partial/\partial
x^{\mu}=1/{x'}^2 {\cal M}_{\mu\nu}(x')\partial/\partial
{x'}_{\nu}$.}

Besides of (3.2a,b) we introduce the following boundary condition for
the fifth coordinate $x_5$  

$${i\over M}{ {\partial}\over { \partial x_5}}\varphi_a (x,x_5) =
\eta_a\varphi_a(x,x_5)+l_a(x,x_5);\ \ \ a=1,2\equiv inr,ext, \eqno(3.4)$$

where
$$\eta_{inr}=\sqrt{|1-m^2/M^2|};\ \ \ \eta_{ext}=\sqrt{1+m^2/M^2}.\eqno(3.5)$$
Acting with  $M^2(i/M\partial/\partial x_5 +\eta_a)$ ¢
on the relation (3.4) we get

$$\Bigl({ {\partial^2}\over { \partial x_5 \partial x^5}} +M^2-m^2\Bigr)
\varphi_{inr}(x,x_5)=
-M^2\Bigl({i\over M} {{\partial}\over { \partial x_5}}+\eta_{inr}\Bigr)
l_{inr}(x,x_5)  \eqno(3.6a)$$
and
$$\Bigl({ {\partial^2}\over { \partial x_5 \partial x^5}} +M^2+m^2\Bigr)
\varphi_{ext}(x,x_5)= -M^2\Bigl({i\over M}{ {\partial}\over {
\partial x_5}}+\eta_{ext}\Bigr) l_{ext}(x,x_5).  \eqno(3.6b)$$
Combining these equations with (3.3a,b) we obtain
$$\Bigl({ {\partial^2}\over { \partial x_{\mu} \partial x^{\mu}}} +m^2\Bigr)
\varphi_{inr}(x,x_5)=
M^2\Bigl({i\over M}{ {\partial}\over { \partial x_5}}+\eta_{inr}\Bigr)
l_{inr}(x,x_5)$$
$$\equiv j_{inr}(x,x_5)\eqno(3.7a)$$

$$\Bigl({ {\partial^2}\over { \partial x_{\mu} \partial x^{\mu}}} +m^2\Bigr)
\varphi_{ext}(x,x_5)=
-M^2\Bigl({i\over M}{ {\partial}\over { \partial x_5}}+\eta_{ext}\Bigr)
l_{ext}(x,x_5)$$
$$\equiv j_{ext}(x,x_5),\eqno(3.7b)$$
where 
$j_{a}(x,x_5)$  are determined via $l_{a}(x,x_5)$.

Solutions of eq.(3.7a,b) determine the solutions of  4D
equation (3.1a) with¨
$$J(x)\equiv J(x,x_5=t_5)=j_{inr}(x,x_5=t_5)+j_{ext}(x,x_5=t_5)
\eqno(3.8)$$

On the other hand the boundary condition (3.4) can be presented in the  
integral form

$$\varphi_{a}(x,x_5)=e^{-iM(x_5-t_5)}\biggl\{\varphi_a(x,x_5=t_5)-
{1\over {\eta_a}}l_a(x,x_5=t_5)\bigl[ e^{2iM\eta_a(x_5-t_5)} -1\bigr]$$
$$
-{1\over {2iM\eta_a}}\int_{t_5}^{x_5} dz_5 j_a(x,z_5) e^{-iM\eta_a(z_5-t_5)}
\bigl[ e^{2iM\eta_a(x_5-t_5)} -e^{2iM\eta_a(z_5-t_5)}\bigr]
\biggr\}. \eqno(3.9)$$
For noninteracting particles, when
  $l_{a}(x,x_5)=0$ and $j_{a}(x,x_5)=0$, equations (3.7a,b), 
(3.3a,b) and (3.4) coincide with the similar
equations and constraints from ref. \cite{K2,K3}.

{\underline {\bf Consistency condition for the 5D equation of motion
 (3.7a,b) and the }}
\newline
{\underline {\bf boundary conditions (3.3a,b) and (3.4):}}\ \ \ \ \ 
Combining eq.(3.7a,b) and (3.6a,b) we find
$$\Bigl({ {\partial^2}\over { \partial x_{\mu} \partial x^{\mu}}}
{ {\partial^2}\over { \partial x_{5} \partial x^{5}}}-
{ {\partial^2}\over { \partial x_{5} \partial x^{5}}}
{ {\partial^2}\over { \partial x_{\mu} \partial x^{\mu}}}\Bigr)
\varphi_a(x,x_5)$$
$$=
\mp\Bigl({ {\partial^2}\over { \partial x_{\mu} \partial x^{\mu}}} \pm
{ {\partial^2}\over { \partial x_{5} \partial x^{5}}}\mp M^2
\Bigr)j_{a}(x,x_5)=0.\eqno(3.10a)$$

According to this relation, 
5D equations of motion (3.7a,b) 
are consistent with the boundary conditions (3.3a,b) and 
(3.4) if
$j_{a}(x,x_5)$ and $l_{a}(x,x_5)$ 
are embedded in hyperboloid  $q^2\pm q_5^2=\pm M^2$, i.e.
in analogue to $\varphi_{a}(x,x_5)$ the operators
$j_{a}(x,x_5)$ and $l_{a}(x,x_5)$ must satisfy the
conditions

$$\biggl({{\partial^2}\over{\partial x^{\mu}\partial x_{\mu}}}+
{{\partial^2}\over{\partial x^5\partial
x_5}}+M^2\biggr)j_{inr}(x,x_5)=0, \eqno(3.10b)$$

$$\biggl({{\partial^2}\over{\partial x^{\mu}\partial x_{\mu}}}-
{{\partial^2}\over{\partial x^5\partial
x_5}}-M^2\biggr)j_{ext}(x,x_5)=0. \eqno(3.10c)$$

Using the conditions (3.3a,b) $\varphi_{a}(x,x_5)$ may be represented
as

$$\varphi_{inr}(x,x_5)={2M\over{(2\pi)^4}}\int d^5q
e^{-iq_5x^5} \delta(q^2+q_5^2-M^2)
[\theta(q^2)\theta(M^2-q^2)+\theta(-q^2)\theta(-M^2-q^2)]$$
$$\Bigl[ e^{-iqx}\varphi^{(+)}_{inr}(q,q_5)+e^{iqx}\varphi^{(-)}_{inr}(q,q_5)
\Bigr],\eqno(3.11a)$$
and
$$\varphi_{ext}(x,x_5)={2M\over{(2\pi)^4}}\int d^5q
e^{-iq_5x^5} \delta(q^2-q_5^2+M^2)
[\theta(q^2)\theta(-M^2+q^2)+\theta(-q^2)\theta(M^2+q^2)]$$
$$\Bigl[ e^{-iqx}\varphi^{(+)}_{ext}(q,q_5)+e^{iqx}\varphi^{(-)}_{ext}(q,q_5)
\Bigr].\eqno(3.11b)$$


From (3.10b,c) we get the same representation for source operator

$$j_{inr,ext}(x,x_5)={2M\over{(2\pi)^4}}\int d^5q
e^{-iq_5x^5} \delta(q^2\pm q_5^2\mp M^2)
[\theta(q^2)\theta(\pm M^2 \mp q^2)+\theta(-q^2)\theta(\mp M^2\mp q^2)]$$
$$\Bigl[ e^{-iqx}j^{(+)}_{inr,ext}(q,q_5)+e^{iqx}j^{(-)}_{inr,ext}(q,q_5)
\Bigr].\eqno(3.12)$$

The same representation is valid  for $l_{inr,ext}(x,x_5)$.

\par
Present formulation has a number of common properties with the  other
5D field-theoretical approaches based on the proper time 
method \cite{IZ,Fanchi,Land,Kubo,Green}, where
 $x_5^2\equiv\tau=x_0^2-{\bf x}^2\equiv x_{\mu}x^{\mu}$.
From this point of view the boundary conditions
(3.4) has a form of an evolution equation.
But $l_a$ as well as $j_a$ are  variation of  Lagrangians 
over the interacted fields, 
i.e. eq.(3.4) can not be treated as an evolution equation.
On the other hand the fifth momentum $q_5$
 is singlevalued determined  via the scale parameter
 $\lambda$  (see table 2). Therefore   
 $x_5$ may be provided  with the  a scale interpretation if we take
$\lambda^{-1}=ln(Mx_5)$.
Unlike  other 5D approaches \cite{Fanchi,Snyder,Yang,Hamilton},
in the present formulation field operators and
the source operators are defined in the 5D space with the invariant forms 
(2.2a,b).


\vspace{0.7cm}

\centerline{\bf{4.\ 5D Lagrangian approach}} \vspace{0.5cm}

\par
The 5D operators
 $\varphi_{inr}(x,x_5)$ and $\varphi_{ext}(x,x_5)$ are independent because
they are defined
in the different domains of $q^2\equiv q_{\mu}q^{\mu}$ and $q_5^2$
 variables.
Therefore the sought 5D Lagrangian ${\cal L}\equiv {\cal L}(x,x_5)$ 
must be constructed using the two sets of the independent fields  
$\varphi_{inr}(x,x_5)$ and $\varphi_{ext}(x,x_5)$.
A complete 5D Lagrangian has a form

 $${\cal L}={\cal L}_0+{\cal L}_{INT}+{\cal L}_c,\eqno(4.1a)$$
where ${\cal L}_0$ stands for the noninteracting part 

$${\cal L}_0=\sum_{a=1,2}\Bigl[
{{\partial\varphi_a(x,x_5)}\over{\partial x_{\mu}}}
{{\partial\varphi^{+}_a(x,x_5)}\over{\partial x^{\mu}}}-m^2
\varphi_a(x,x_5)\varphi^{+}_a(x,x_5)\Bigr],\eqno(4.1b)$$

$${\cal L}_{INT}\equiv{\cal L}_{INT}
(\varphi_a,\varphi^{+}_a, {{\partial\varphi_a}/{\partial
x_{\mu}}}, {{\partial\varphi^{+}}_a/{\partial x^{\mu}}};
{{\partial\varphi_a}/{\partial x_{5}}},
{{\partial\varphi^{+}_a}/{\partial x^{5}}})$$
is the interacting part of Lagrangian and
 ${\cal L}_c$ generates  the constraint (3.4)

$${\cal L}_c=M^2\sum_{a=1,2}
|{i\over M}{{\partial\varphi_a}\over{\partial x_{5}}}-\eta_{a}\varphi_a
-l_a(x,x_5)|^2,\eqno(4.1c)$$
where $l_a(x,x_5)\equiv l_a(\varphi_a,\varphi^{+}_a,
{{\partial\varphi_a}/{\partial x_{\mu}}},
{{\partial\varphi^{+}_a}/{\partial x^{\mu}}};
{{\partial\varphi_a}/{\partial x_{5}}},
{{\partial\varphi^{+}_a}/{\partial x^{5}}})$ 
is in response to the interaction in the constraint (3.4).
\par
Next we consider   action
$${\cal S}(x_5)=\int d^4x {\cal L}(x,x_5)\eqno(4.2)$$
and its  variation 
under the conformal transformations (1.1a)-(1.1e)
$$\delta q_{\mu}=\delta h_{\mu}+\delta \Lambda_{\mu\nu}q^{\nu}+
\delta \lambda q_{\mu}+(q^2\delta {\hbar}_{\mu}- 2q^{\nu}\delta
{\hbar}_{\nu} q_{\mu})/M^2,\eqno(4.3a)$$ 
where  $\delta h_{\mu}(\delta
\hbar_{\mu})$, $\delta \Lambda_{\mu\nu}q^{\nu}=-\delta
\Lambda_{\nu\mu}q^{\nu}$ ¨ $\delta \lambda$ 
stands for the infinitesimal parameters of the corresponding transformations. 

Translation $q_{\mu}'=q_{\mu}+h_{\mu}$ does not change
 $x_{\mu}$. Therefore the variation of coordinates
 $\delta x_{\mu}=x_{\mu}'-x_{\mu}$, generated by variation of four momenta
 $q_{\mu}$  (4.3), includes only the rotation and the scale transformations

$$\delta x_{\mu}=\delta \Lambda_{\mu\nu}^{-1}x^{\nu}-
\delta \lambda x_{\mu}\eqno(4.3b)$$

In the considered formulation
$x_5$ is independent variable. Therefore we take

$${ {\delta x_5}\over {\delta x_{\mu}} }=
\delta {{d x_5}\over {d x_{\mu}} }=0\eqno(4.4)$$
which is consistent with our choice of action (4.2).

Now we have
$$\delta {\cal S}(x_5)=\sum_{a=1,2}\biggl\{\int d^4 x\biggl[
{{\partial{\cal L}}\over{\partial \varphi^{+}_a(x,x_5)}}
-{{\partial}\over{\partial x_{\mu}}} \Bigl(
 {{\partial{\cal L}}\over{\partial
[{{\partial \varphi^{+}_a(x,x_5) }/{\partial x^{\mu} ] }} }}
    \Bigr)\biggr]{\overline \delta}\varphi_a^{+}(x,x_5)$$

$$+\int d^4 x {{d}\over{d x_{\mu}}} \Bigl[
{{\partial{\cal L}}\over{\partial \ [
{{\partial \varphi^{+}_a(x,x_5) }/{\partial x^{\mu} ] }} }}
{\overline  \delta}\varphi_a^{+}(x,x_5) + {\cal L}(x,x_5) \delta x^{\mu}
\Bigr]\biggr\}$$

$$+\int d^4 x  \Bigl[
{{\partial{\cal L}}\over{\partial  [
{{\partial \varphi^{+}_a(x,x_5) }/{\partial x^{5}   ] }} }}
 {\overline  \delta}\bigl(
{{\partial}\over{\partial x_{5}}}\varphi_a^{+}(x,x_5)\bigr) +
{{d{\cal L} }\over{d x_{5}}}\delta x^{5}
\Bigr]$$
$$\ +\ hermitian\ conjugate\eqno(4.5)$$
where  ${\overline  \delta}$ denotes a variation of form of the corresponding 
expression. Substituting
 ${{d{\cal L} }/{d x_{5}}}$ in (4.5)  we obtain
$$\delta {\cal S}(x_5)=\sum_{a=1,2}\biggl\{\int d^4 x\biggl[
{{\partial{\cal L}}\over{\partial \varphi^{+}_a(x,x_5)}}
-{{\partial}\over{\partial x_{\mu}}} \Bigl(
 {{\partial{\cal L}}\over{\partial
[{{\partial \varphi^{+}_a(x,x_5) }/{\partial x^{\mu} ] }} }}
    \Bigr)\biggr]\biggl[{\overline \delta}\varphi_a^{+}(x,x_5)+
{{\partial \varphi^{+}_a(x,x_5) }\over{\partial x^{5} }}\delta x_5
\biggr]\eqno(4.6a)$$
$$+\int d^4 x {{d}\over{d x_{\mu}}} \Bigl[
{{\partial{\cal L}}\over{\partial \ [
{{\partial \varphi^{+}_a(x,x_5) }/{\partial x^{\mu} ] }} }}
{\overline  \delta}\varphi_a^{+}(x,x_5) + {\cal L}(x,x_5) \delta x^{\mu}
\Bigr]\eqno(4.6b)$$
$$+\int d^4 x
{{\partial{\cal L}}\over{\partial  [
{{\partial \varphi^{+}_a(x,x_5) }/{\partial x^{5} ] }} }}
\Bigl[ {\overline  \delta}\bigl(
{{\partial}\over{\partial x_{5}}}\varphi_a^{+}(x,x_5)\bigr) +
{{\partial^2 \varphi^{+}_a(x,x_5) }\over{\partial {x_{5}}^2 }}\delta x_5
\Bigr]\eqno(4.6c)$$
$$+\int d^4 x {{d}\over{d x_{\mu}}}\Bigl[
{{\partial{\cal L}}\over{\partial \ [ {{\partial
\varphi^{+}_a(x,x_5) }/{\partial x^5 ] }} }} {{\partial
\varphi_a^{+}(x,x_5)}\over{\partial x_{\mu} }}\delta x_5
\Bigr]\biggr\}\eqno(4.6d)$$
$$\ +\ hermitian\ conjugate$$

In order to get $\delta {\cal S}(x_5)=0$  we shall suppose
that every term of eq.(4.6) vanishes. Then 
for the every term separately we obtain the following equations:

\begin{itemize}

\item[$1.$]
The first term (4.6a) represents the equation of motion for
$\varphi_a^{+}(x,x_5)$ and $\varphi_a^{+}(x,x_5)$
$${{\partial{\cal L}}\over{\partial \varphi^{+}_a(x,x_5)}}
={{d}\over{d x_{\mu}}} \Bigl(
 {{\partial{\cal L}}\over{\partial
[{{\partial \varphi^{+}_a(x,x_5) }/{\partial x^{\mu} ] }} }}\Bigr);\ \ \
{{\partial{\cal L}}\over{\partial \varphi_a(x,x_5)}}
={{d}\over{d x_{\mu}}} \Bigl(
 {{\partial{\cal L}}\over{\partial
[{{\partial \varphi_a(x,x_5) }/{\partial x^{\mu} ] }}
}}\Bigr),\eqno(4.7a)$$

or
$$\Bigl({{\partial^2}\over{\partial x_{\mu} \partial x^{\mu}}}
+m_0^2\Bigr)\varphi_a(x,x_5)= {{\partial{\cal
L}_{INT}}\over{\partial \varphi^{+}_a(x,x_5)}} -{{d}\over{d
x_{\mu}}} \Bigl(
 {{\partial{\cal L}_{INT}}\over{\partial
[{{\partial \varphi^{+}_a(x,x_5) }/{\partial x^{\mu}] }} }}
    \Bigr)\equiv j_a(x,x_5)\eqno(4.7b)$$
which coincides with (3.7a,b).

\item[$2.$]
The next term (4.6b) relates to the 4D current conservation condition

$${\cal J}^{\mu}(x)=\sum_{a=1,2}{\cal J}^{\mu}_a(x,x_5=t_5),\eqno(4.11a)$$
where
$${\cal J}^{\mu}_a(x,x_5)={{\partial{\cal L}}\over{\partial \ [
{{\partial \varphi^{+}_a(x,x_5) }/{\partial x_{\mu} ] }} }}
{\overline  \delta}\varphi_a^{+}(x,x_5) + {\cal L}(x,x_5) \delta
x^{\mu}$$
$$\ +\ hermitian\ conjugate.\eqno(4.7b)$$

\item[$3.$]
Third term  (4.6c) contains
 $\partial\varphi_a(x,x_5) /\partial x^{5}$. This field may be treated
as independent due to fifth degrees of freedom \cite{K2,K3}.
Therefore we can introduce  a new kind of fields
$$\chi_a(x,x_5)={i\over M}
{{\partial\varphi_a(x,x_5) }\over{\partial x_{5} }}.\eqno(4.8)$$
Using the variation principle and  the
independence of the fields $\chi_a(x,x_5)$ we get 
$$
{{\partial{\cal L}}\over{\partial  [ {{\partial
\varphi^{+}_a(x,x_5) }/{\partial x^{5}   ] }} }} =
{{\partial{\cal
L}}\over{\partial  [ {{\partial \varphi_a(x,x_5) }/{\partial x^{5}
] }} }}=
{{\partial{\cal
L}_c}\over{\partial  [ {{\partial \varphi_a(x,x_5) }/{\partial x^{5}
] }} }}=0 \eqno(4.9)$$ 
which implies

$$\chi_a-\eta_{a}\varphi_a-l_a(x,x_5)=-1/M^2
{{\partial{\cal L}_{INT} }\over{\partial \chi^{+}_a(x,x_{5})}}$$
$$+{{\partial l^{+}_a }\over{\partial \chi^{+}_a(x,x_{5})}}
\Bigl( \chi_a-\eta_{a}\varphi_a-l_a(x,x_5)\Bigr) +{{\partial l_a
}\over{\partial \chi^{+}_a(x,x_{5})}} \Bigl(
\chi^{+}_a-\eta_{a}\varphi^{+}_a-l_a(x,x_5)^{+}\Bigr).
\eqno(4.10)$$

Afterwards we restrict our formulation
with such 
${\cal L}_{INT}$ which are independent on  ${{\partial\varphi_a}/{\partial
x_{5}}}$ and  ${{\partial\varphi^{+}}/{\partial x^{5}}}$.
Then instead of (4.10) we get

$$\chi_a(x,x_5)-\eta_{a}\varphi_a(x,x_5)-l_a(x,x_5)=0,\eqno(4.11)$$
which coincides with  (3.4).

 Combining (3.2a,b) and (4.11) we get the connections between
$l_a(x,x_5)$ and $j_a(x,x_5)$

$$j_{a}(x,x_5)=(-1)^{a-1}
M^2\Bigl({i\over M}{ {\partial}\over { \partial
x_5}}+\eta_{a}\Bigr) l_{a}(x,x_5)\eqno(4.12)$$
which was presented in eq. (3.7a,b).

\item[$4.$]
The fourth term  (4.6d) contains the current operator
$$ J^{\mu}_a(x,x_5)=
{{\partial{\cal L}}\over{\partial \ [ {{\partial
\varphi^{+}_a(x,x_5) }/{\partial x^{\mu} ] }} }} {{\partial
\varphi_a^{+}(x,x_5)}\over{\partial x_{5} }} \delta x_5$$
$$\ +\ hermitian\ conjugate\eqno(4.13).$$

in the asymptotic region, where $j_a=0$ and $l_a=0$,
 $ J^{\mu}_a(x,x_5)$ has the same form as the electro-magnetic 
current operator

$$ J^{\mu}_a(x,x_5)=-i\eta_a\delta x_5\biggl\{
{{\partial{\cal L}_0}\over{\partial \ [
{{\partial \varphi_a(x,x_5) }/{\partial x^{\mu} ] }} }}
 \varphi_a(x,x_5)$$
$$-
{{\partial{\cal L}_0}\over{\partial \ [
{{\partial \varphi^{+}_a(x,x_5) }/{\partial x^{\mu} ] }} }}
 \varphi_a^{+}(x,x_5)
\biggr\}\eqno(4.14)$$

 and $ {{d}/{d x_{\mu}}}J^{\mu}_a(x,x_5)=0$.
For the interacting fields expression (4.13) vanishes if we require 
that $\delta x_5=0.$

\end{itemize}

Thus we have derived the equation of motion (3.7a,b), constraint
(3.4) and the expression for the conserved currents (4.7b) and (4.13)
using the variation principle. Combining these equations one can verify, that
$d {\cal S}(x_5) / dx_5=0,$ i.e. ${\cal S}(x_5)$ (4.2) is not
dependent on $x_5$. In particular, using the 5D equations of motion
(4.7a,b) we get $d {\cal S}(x_5) /d x_5
=\int d^4x d/d x_{\mu} \Bigl\{ \partial {\cal L}
/\partial[{{\partial \varphi^{+}_a(x,x_5)} /{\partial x^{\mu} }}]
\partial \varphi_a^+(x,x_5)/\partial x_{5}+\ í.\ á.\Bigr\}$
which vanishes according to fifth  current conservation condition 
(4.13).

\newpage


\centerline{\bf{5.\ Construction  interaction part of Lagrangian 
  ${\cal L}_{INT}$  via constrain (3.4).}}
\vspace{0.75cm}

\par
In present formulation all field operators 
in equation of motion (3.7a,b), in constrain (3.4) and in Lagrangian
(4.1a,b,c) are 
defined on the surface $q^2\pm q_5^2=\pm M^2$.
But the product of $\varphi_a(x,x_5)$ is not on  
$q^2\pm q_5^2=\pm M^2$ shell. Therefore in this section we 
consider
off $q^2\pm q_5^2=\pm M^2$ shell representations
of equation of motion (3.7a,b) with corresponding off shell
constrain (3.4) and off shell Lagrangian.
Afterwards using the simple projection procedure, we put 
these Lagrangians, equation of motions with related constrains on 
$q^2\pm q_5^2=\pm M^2$ surface.
We introduce off   $q^2\pm q_5^2\mp M^2$ shell operator
 ${\widetilde \varphi_a}(x,x_5)$  as follows

$$\varphi_{a}(x,x_5)=\int d^5y {\widetilde \varphi_a}(x-y,x_5-y_5)
D_{a}(y,y_5),\eqno(5.1)$$
where

$$\varphi_{a=1}(x,x_5)={2M\over{(2\pi)^4}}\int d^5q
e^{-iq_5x^5} 
[\theta(q^2)\theta(M^2-q^2)+\theta(-q^2)\theta(-M^2-q^2)]$$
$$\Bigl[ e^{-iqx}\varphi^{(+)}_{1}(q,q_5)+e^{iqx}\varphi^{(-)}_{1}(q,q_5)
\Bigr],\eqno(5.2a)$$
and
$$\varphi_{a=2}(x,x_5)={2M\over{(2\pi)^4}}\int d^5q
e^{-iq_5x^5} 
[\theta(q^2)\theta(-M^2+q^2)+\theta(-q^2)\theta(M^2+q^2)]$$
$$\Bigl[ e^{-iqx}\varphi^{(+)}_{2}(q,q_5)+e^{iqx}\varphi^{(-)}_{2}(q,q_5)
\Bigr].\eqno(5.2b)$$

and

$$D_a(x,x_5)={1\over{(2\pi)^5}}\int d^5q
e^{-iqx-iq_5x^5}\delta(q^2\pm q_5^2\mp M^2).\eqno(5.3)$$
Substituting (5.2a,b) and (5.3) into (5.1) we obtain expressions (3.11a,b)  
after Fourier transforms.

In the same manner as for ${\widetilde \varphi_a}(x,x_5)$ (5.1)
we introduce

$$j_{a}(x,x_5)=\int d^5y {\widetilde j_a}(x-y,x_5-y_5)
D_{a}(y,y_5),\eqno(5.4)$$

$$l_{a}(x,x_5)=\int d^5y {\widetilde l_a}(x-y,x_5-y_5)
D_{a}(y,y_5),\eqno(5.5)$$

$${\cal L}_{a}(x,x_5)=\int d^5y {\widetilde {\cal L}_a}(x-y,x_5-y_5)
D_{a}(y,y_5),\eqno(5.6)$$

Then  from the equations of motion
$$\biggl[ { {\partial^2}\over{ \partial x_{\mu}\partial x^{\mu}}}
 +m^2\biggr]{\widetilde \varphi_a}(x-y,x_5-y_5)= {\widetilde
j_a}(x-y,x_5-y_5),\eqno(5.7)$$
we obtain the equations of motion (3.7a,b)
using integration over $y,y_5$ variables according to (5.1) and (5.4).

In the same way as eq.(4.7a,b) with the constraint (4.11)
we can  derive equation of motion (5.7) and the off shell constraint
 for the off $q^2\pm q_5^2\mp M^2$ shell Lagrangian

$${\widetilde{\cal L}_a}=({\widetilde{\cal L}_a})_0+
({\widetilde{\cal L}_a})_{INT}+({\widetilde{\cal L}_a})_c,\eqno(5.8a)$$
where $({\widetilde{\cal L}_a})_0$ stands for the noninteracting part,
$({\widetilde{\cal L}_a})_{INT}$ is  the interaction part
and $({\widetilde{\cal L}_a})_c$  generate the constraint

$${\widetilde\chi_a}(x,x_5)-\eta_{a}{\widetilde\varphi_a}(x,x_5)-
{\widetilde l_a}(x,x_5)=0\eqno(5.9a)$$
and have the form
$$({\widetilde{\cal L}})_c=M^2\sum_{a=1,2}
|{i\over M}{{\partial{\widetilde\varphi}_a}\over{\partial x_{5}}}
-\eta_{a}{\widetilde\varphi}_a
-{\widetilde l}_a(x,x_5)|^2.\eqno(5.9b)$$
Thus relations (5.1), (5.4), (5.5) and (5.6) 
enables us to obtain the straightforward off shell
representations of eq.(3.7a,b) and the constraint (3.4) using the similar
as (4.1a,b,c) Lagrangian but with off 
$q^2\pm q_5^2\mp M^2$ shell operators ${\widetilde\varphi}_a$.
Now we consider some example of
 ${\widetilde l_a}(x,x_5)$ and the corresponding interaction Lagrangians:

\vspace{0.7cm}
\par
{\underline {\bf $\varphi^4$ model:}}
The simplest 
 ${\widetilde l_a}$ which is not dependent on
 ${\widetilde\chi_a}(x,x_5)\equiv
i/M\ \partial {\widetilde\varphi_a} (x,x_5)/\partial x_5$ is
$${\widetilde l_a}=g_a{\widetilde\varphi_a}^2,\eqno(5.10)$$

Using the constraint (5.9a)  we get

$${i\over M}{{\partial{\widetilde\varphi_a}(x,x_5) }\over{\partial x_{5} }}
=\eta_{a}\
{\widetilde\varphi_a}(x,x_5)+{g_a}{\widetilde\varphi_a}^2(x,x_5).
\eqno(5.11)$$

$${\widetilde j_{a}}(x,x_5)=\partial {\widetilde{\cal L}_{INT}}/\partial
{\widetilde \varphi^{+}_{a}}(x,x_5)
=(-1)^{a-1}M^2\Bigl({i\over M}{{\partial}\over{\partial x_{5} }}
+\eta_{a} \Bigr){\widetilde l_{a}}(x,x_5)$$
$$=(-1)^{a-1}M^2\Bigl(3g_a  \eta_{a}{\widetilde\varphi_a}^2(x,x_5)+
2g_a^2{\widetilde\varphi}^3(x,x_5)\Bigr).\eqno(5.12)$$ 
The corresponding equation of motion can be derived using the following
Lagrangians

$${\widetilde{\cal L}}={1\over 2}\sum_{a=1,2}\Bigl[
{{\partial {\widetilde\varphi_a}}\over{\partial x_{\mu}}}
{{\partial{\widetilde\varphi_a}}\over{\partial x^{\mu}}}-m^2_a
{\widetilde\varphi_a}^2\Bigr]+ M^2\sum_{a=1,2} |\ {\widetilde\chi_a}-\eta_{a}
{\widetilde\varphi_a} -{g_a}{\widetilde\varphi_a}^2\ |^2+ 
{\widetilde {\cal L}}_{INT},\eqno(5.13)$$ 
where 

$$({\widetilde{\cal L}_a})_{INT}(x,x_5)
=(-1)^{a-1}M^2\Bigl(  {g_a} \eta_{a} {\widetilde\varphi}_a^3 +
{{g_a^2}\over 2} {{\widetilde \varphi}_a^4}\Bigr)\eqno(5.14)$$

Lagrangian  (5.13) has  the following attractive properties
\begin{itemize}

\item[${\bf I.}$]
The considered model is renormalizable, because
${\widetilde{\cal L}}_{INT}$ and ${\widetilde{\cal L}}_{á}$ contains  
${\widetilde \varphi}_a$ in the third and in the
fourth power.

\item[${\bf II.}$]

$({\widetilde{\cal L}_{inr}})_{INT}$ $(a=1)$ and 
$({\widetilde{\cal L}_{ext}})_{INT}$ $(a=2)$ have the opposite sign.

\item[${\bf III.}$]
The Lagrangian  (5.14) has a local minimum at
 $-2\eta_{a}/ g_a$ and a local maximum at  $-{3\over2}\eta_{a}/ g_a$.

\end{itemize}

\vspace{0.7cm}
\par
{\underline {\bf
Nonlinear $\sigma$ model:}} \ \ \ \ \
Here we have $\pi$-meson fields $\pi^{\pm}$, $\pi^{0}$
instead of $\varphi$. We choose   $l_a$ 
depending on the auxiliary fields   $\chi$ 
$${\widetilde l_a}^{\alpha}={1\over {4f_{\pi}^2}}({\widetilde\chi_a}^{\gamma}
 {{\widetilde\chi}^{\gamma}_a})
{\widetilde\pi}_a^{\alpha}\equiv {1\over
{4f_{\pi}^2}}{\widetilde\chi}_a^2
{\widetilde\pi}_a^{\alpha},\eqno(5.15)$$ 
where we have used the well known isospin redefinition 
of the $\pi$-meson fields
$\pi^{\pm}\equiv 1/2(\pi^1\pm i \pi^2);\ \ \ \pi^{0}\equiv \pi^3$;
$\alpha,\beta,\gamma=1,2,3$, $f_{\pi}=93MeV$ is the $\pi$-meson decay
 constant and Lagrangian   
  is  choosing in the form
$${\widetilde{\cal L}}=\sum_{a=1,2}\biggl( {1\over 2}
{{\partial}\over{\partial{x_\mu}}} {\widetilde\pi}_a^{\alpha}
{{\partial}\over{\partial{x^\mu}}} {{\widetilde\pi}^{\alpha}_a}
+M^2\Bigl[
{\widetilde\chi_a}^{\alpha}-{\widetilde\pi_a}^{\alpha}-
{1\over {4f_{\pi}^2}}{\widetilde\chi_a}^2
{\widetilde\pi_a}^{\alpha}
\Bigr]^2\biggr) +{\widetilde{\cal L}}_{chir}+
{\widetilde{\cal L}}_{INT},\eqno(5.16)$$

where the second term generates the constraint between the
auxiliary field
  ${\widetilde\chi_a}^{\alpha}(x,x_5)=
{i/ M}{{\partial{\widetilde\pi_a}^{\alpha}(x,x_5) }/{\partial x_{5}}}$ 
and the $\pi$ meson field ${\widetilde \pi_a}^{\alpha}$

$${\widetilde\chi_a}^{\alpha}-{\widetilde\pi_a}^{\alpha}-
{1\over {4f_{\pi}^2}}{\widetilde\chi_a}^2
{\widetilde\pi_a}^{\alpha}=0.\eqno(5.17a)$$ 

This constraint coincides with the relation between the 
$\pi$ meson field and the
interpolating field in the nonlinear
$\sigma$-model \cite{Wei,Alf}
$${\widetilde\pi_a}^{\alpha}={1\over{1+
{ {{\widetilde\chi_a}^2 }\over{4f_{\pi}^2} } } } 
{\widetilde \chi_a}^{\alpha}.\eqno(5.17b)$$

Third term of (5.16)  ${\widetilde{\cal L}}_{chir}$ reproduces the  
constraint between $\pi$-meson fields and the auxiliary 
 $\sigma$-meson fields
$${\widetilde\pi_a}^2+{\widetilde\sigma}_a^2=f_{\pi}^2
\eqno(5.18)$$

and correspondingly

$$({\widetilde{\cal L}}_a)_{chir}(x,x_5)=
\Bigl({\widetilde\pi_a}^2+{\widetilde\sigma}_a^2-f_{\pi}^2\Bigr)^2.
\eqno(5.19)$$

In the usual  $\sigma$ model
the chiral symmetry is weakly broken with the additional Lagrangian
${\cal L}'=-f_{\pi}m_{\pi}^2\sigma$. In the considered model
the chiral symmetry breaking terms arise in  
$[{\widetilde{\cal L}}_A]_{INT}$ as the result of connection between
${\widetilde{l}_{a}}$ (5.15) with $j_a$ and
$[{\widetilde{\cal L}}_a]_{INT}$ 
. Thus the source operator ${\widetilde j_{a}}^{\alpha}$  
is defined via operator (5.15) as

$${\widetilde j_{a}}^{\alpha}(x,x_5)=
\partial ({\widetilde{\cal L}_a})_{INT}/
\partial {\widetilde\pi^{\alpha}}_{a}(x,x_5)
=(-1)^{a-1}M^2\Bigl({i\over M}{{\partial}\over{\partial x_{5} }}+1\Bigr)
{\widetilde l_{a}}^{\alpha}(x,x_5)$$

$$=(-1)^{a-1} M^2{{(f_{\pi}+{\widetilde\sigma_a})}
\over{{\widetilde\sigma_a}} } \Bigl[ 1+ f_{\pi}
{{(3f_{\pi}-{\widetilde\sigma_a})}
\over{(f_{\pi}-{\widetilde\sigma_a})^2} } \Bigr]
{\widetilde\pi_a}^{\alpha} .\eqno(5.20)$$ 

The corresponding Lagrangian is
$${\widetilde{\cal L}}_{INT}=-\sum_{a=1,2}(-1)^{a-1}M^2
\Bigl(f_{\pi}{\widetilde\sigma_a} +{1\over 2}{\widetilde\sigma_a}^2
+f_{\pi}  {{(f_{\pi}+{\widetilde\sigma_a})^2}
\over{(f_{\pi}-{\widetilde\sigma_a})} } \Bigr)
\eqno(5.21a)$$
which after expansion in  $\pi_a^2$ power series
takes the form
$${\widetilde{\cal L}}_{INT}=-\sum_{a=1,2}(-1)^{a-1}M^2
\Bigl(-{9\over 2}f_{\pi}^2+ 8{{f_{\pi}^4}\over{ {\widetilde
\pi_a}^2} }-{\widetilde \pi_a}^2 -{1\over{4f_{\pi}^2} }{\widetilde
\pi_a}^4-...\Bigr) \eqno(5.21b)$$
Afterwards  ${\widetilde{\cal L}}_{INT}$    obtains 
the real $\pi$ meson mass term for the  $ext=a\equiv 2$ 
if $M$ is fixed as
$$M={{m_{\pi}}\over{\sqrt{2} }}.\eqno(5.22)$$
and for the internal $\pi$ meson field ${\widetilde \pi}_{a=1}$
in (5.21b) appear only negative $m_{\pi}^2$, i.e. ${\widetilde\pi}_{a=1}$
remains to be massless.

Thus the simple form of ${\widetilde l_{a}}$ (5.15) 
and constraint (5.17b) allows us
to interpret ${\widetilde\chi}=i/M\partial/\partial x_5{\widetilde\pi}$ as 
the auxiliary interpolating pion field  according to the
nonlinear $\sigma$-model \cite{Wei,Alf}.
From the other hand the same ${\widetilde l_{a}}$ (5.15) determines the source
 operator (5.20) and corresponding Lagrangian (5.21a) which
consists from
the usual chiral symmetry breaking term  of  $\sigma$ models 
$-m_{\pi}^2 f_{\pi}{\widetilde\sigma_a}$
and other terms which break chiral symmetry more strongly.
The 4D $\pi$-meson field operator
${\widetilde\pi}^{\alpha}(x)={\widetilde\pi}_{a=1}^{\alpha}(x,x_5=t_5) 
+{\widetilde\pi}_{a=2}^{\alpha}(x,x_5=t_5);$\ \ \ $(t_5=0\ or\ \sqrt{x_0^2-
{\bf x}^2})$  is similar to
 the  pion field of the nonlinear $\sigma$ model
 in the region $q^2>M^2=m_{\pi}^2/2$, where 
${\widetilde\pi}^{\alpha}(x)={\widetilde\pi}_{a=2}^{\alpha}(x,x_5=t_5).$
In the region $0<q^2<M^2=m_{\pi}^2/2$, where
${\widetilde\pi}_a^{\alpha}(x)={\widetilde\pi}_{a=1}^{\alpha}(x,x_5=t_5)$,
$({{\widetilde{\cal L}}_{a=1}})_{INT}$ has the opposite 
to $({{\widetilde{\cal L}}_{a=2}})_{INT}$ sign.  
In the limit $m_{\pi}\to 0$ (i. e. $M^2\to 0$,) the 
above  Lagrangian transforms into the free Lagrangian for the massless
pion. Note that the chiral symmetry breaking mechanism  
allowed us to fix the scale parameter $M$ of the conformal
transformation group.

The considered 5D nonlinear $\sigma$ model
as well as the ordinary 4D nonlinear $\sigma$ model
is unrenormalizable.
In order to make this model renormalizable we
must introduce  the auxiliary scalar  $\sigma$ field according to 
the constraint (5.18) and the other auxiliary field 
which compensates the $1/\pi^2$ nonlinearity of the last term in (5.21a).
On the other hand inversion procedure transforms the internal region
with $q^2<M^2$
and the corresponding inversion-invariant(anti-invariant) field operators
(1.25a,b)
into  external region $q^2>M^2$ with the same operators (1.25a,b).
Therefore we might hope that this approach will lead to 
specific regularization in the ultraviolet region, where $q^2>>M^2$.


\newpage

\centerline{\bf{6.\ Models with the gauge transformations}}
\vspace{0.75cm}

\par
{\underline {\bf 
 Gauge transformation in the 4D and 5D coordinate space.}

Here we shall treat gauge transformations $q_{\mu}'=q_{\mu}-eA_{\mu}(q)$
as the special form of 4D momentum translation
which can be performed in the same manner as the 
usual 4D translation
$q_{\mu}'=q_{\mu}+h_{\mu}$  (1.1a) 
in the 6D space with the invariant form   $\kappa_A\kappa^A=0$.
In analogue to  eq.(2.5), 
 translations $q_{\mu}'=q_{\mu}-eA_{\mu}(q)$ imply the following
transformations of the 6D variables
$$\kappa_{\mu}'=\kappa_{\mu}-ea_{\mu}(\kappa_A)\kappa_{+}\eqno(6.1a)$$
 with only four auxiliary fields $a_{\mu}(\kappa_A)$ and
$$ \kappa_{+}'=\kappa_{+};\ \ \ 
\kappa_{-}'=\kappa_{-}-
e/M^2\Bigl(
a_{\nu}(\kappa_A)\kappa^{\nu}+\kappa^{\nu}a_{\nu}(\kappa_A)\Bigr)+
e^2/M^2\kappa_+a_{\nu}(\kappa_A)a^{\nu}(\kappa_A)\eqno(6.1b)$$
 Afterwards we get
$${q^2}'=q^2-
e\Bigl({A_a}_{\nu}(q,q_5)q^{\nu}+q^{\nu}{A_a}_{\nu}(q,q_5)\Bigr)+
e^2{A_a}_{\nu}(q,q_5){A_a}^{\nu}(q,q_5)\eqno(6.1c)$$ and
$${q^2_5}'=q^2_5\mp
e\Bigl({A_a}_{\nu}(q,q_5)q^{\nu}+q^{\nu}{A_a}_{\nu}(q,q_5)\Bigr)+
e^2{A}_{a\nu}(q,q_5){A}_a^{\nu}(q,q_5),\eqno(6.1d)$$ where 
the sign $-$ corresponds to
  $q^2+q_5^2=M^2$ and the sign $+$ relates
to  $q^2-q_5^2=-M^2$.
${A_a}_{\nu}(q,q_5)$ is constructed by
$a_{\nu}(\kappa_A)$ in the same way as in eq.(2.14a,b). 
As a result of these gauge 
transformations ´we get ${q^2}'\pm{q_5^2}'={q^2}\pm{q_5^2}$
\footnotemark. 

\par
\footnotetext{In other 5D and 6D formulations with gauge 
 transformations usually were introduced five 
$\Bigl( A_{\nu}(x_{\mu},x_5),A_{5}(x_{\mu},x_5)\Bigr)$
or six $\Bigl( a_{A}(\xi_{\mu},\xi_5,,\xi_6),\Bigr)$
auxiliary gauge fields \cite{Kas,Fanchi,K3}. Such type gauge transformations
violates invariance of 6D ($\kappa_A\kappa^A=0$) or 5D  
(${q^2}\pm{q_5^2}=\pm M^2$)
forms, i.e. these gauge transformations destroy 
the necessary condition .of the conformal transformations.}

We can derive equations of motions 
for the off  $q^2\pm {q_5}^2=\pm M^2$ shell operator   
${\widetilde\varphi}_{a}(x,x_5)$
using gauge transformations $\partial/\partial x'_{\mu}=
\partial/\partial x_{\mu}+ie{\widetilde A}_{a\mu}(x,x_5)$
in the Klein-Gordon equations .
$\Bigl({ {\partial^2}/ { \partial x_{\mu} \partial x^{\mu}}} +m^2\Bigr)
{\widetilde\varphi}_{a}(x,x_5)=0.$
In particular we get

$$\Bigl({ {\partial^2}\over { \partial x_{\mu} \partial x^{\mu}}} +m^2\Bigr)
{\widetilde\varphi}_{a}(x,x_5)= {\widetilde j}_{a}(x,x_5),\eqno(6.2a)$$
where 
$${\widetilde j}_{a}(x,x_5)=\Bigl(ie {{\partial}\over{\partial x_{\mu}
} }{\widetilde A}_{a\mu}(x,x_5)+ie{\widetilde A}_{a\mu}(x,x_5)
{{\partial}\over{\partial
x_{\mu}}}-e^2{\widetilde A}_{a\mu}(x,x_5){\widetilde A_a}^{\mu}
(x,x_5) \Bigr){\widetilde \varphi}_a(x,x_5).\eqno(6.2b)$$ 
The on $q^2\pm {q_5}^2=\pm M^2$ shell operators 
${\varphi}_{a}(x,x_5)$ and $j_{a}(x,x_5)$ are determined via 
${\widetilde \varphi}_{a}(x,x_5)$ and ${\widetilde j}_{a}(x,x_5)$
according to (5.1) and (5.4).
Next we can find ${\widetilde l}_a(x,x_5)$ via ${\widetilde j}_a(x,x_5)$
as

$${\widetilde l}_a(q,q_5)=(-1)^{a-1}{{{\widetilde j}_a(q,q_5)}
\over {M(q_5+M\eta_a)}}\eqno(6.3a)$$
and reproduce the corresponding constraint
$${i\over M}{ {\partial}\over { \partial x_5}}{\widetilde \varphi}_{a}(x,x_5) =
\eta_a{\widetilde \varphi}_{a}(x,x_5)+{\widetilde l}_{a}(x,x_5).
\eqno(6.3b)$$
In addition we can build the 
off $q^2\pm {q_5}^2=\pm M^2$ shell Lagrangian from the source operator 
(6.2b)

$${\widetilde {\cal L}}=\sum_{a=1,2}\Bigl[
{{\partial{\widetilde\varphi}_a(x,x_5)}\over{\partial x_{\mu}}}
{{\partial{\widetilde\varphi}^{+}_a(x,x_5)}\over{\partial
x^{\mu}}}-m^2
{\widetilde\varphi}_a(x,x_5){\widetilde\varphi}^{+}_a(x,x_5)+ M^2
|{i\over M}{{\partial{\widetilde\varphi}_a}\over{\partial
x_{5}}}-\eta_{a}{\widetilde\varphi}_a -{\widetilde
l}_a(x,x_5)|^2\Bigr]$$
$$+\sum_{a=1,2}\biggl[-ie{\widetilde\varphi}_a(x,x_5)
{{ \stackrel{\longleftrightarrow}{\partial}}\over{\partial x_{\mu}} }
{\widetilde \varphi}^{+}_a(x,x_5) {\widetilde
A_a}^{\mu}(x,x_5) +e^2{\widetilde A}_{a\mu}(x,x_5)
{\widetilde A_a}^{\mu}(x,x_5){\widetilde
\varphi}_a(x,x_5){\widetilde\varphi}^{+}_a(x,x_5)
\biggr].\eqno(6.4)$$

Finally we shall build  4D equation of motion 
$$\Bigl( { {\partial^2}\over { \partial x_{\mu} \partial x^{\mu}} }
+m^2\Bigr)\Phi(x)= J(x),\eqno(6.5a)$$

where   $\Phi(x)$ and $J(x)$ consists from the two parts

$$J(x)= j_{a=1}(x,t_5)+j_{a=2}(x,t_5),\ \ \ \ \ 
\Phi(x)= {\phi}_{a=1}(x,t_5)+{\phi}_{a=2}(x,t_5)\eqno(6.5b)$$ 
with arbitrary $M$ and
$t_5=\tau=\sqrt{x_o^2-{\bf x}^2}$ or $t_5=0$. 

It is important to note, that source operator (6.5b) in 4D equation 
(6.5a) differs from the
ordinary source operator in quantum electrodynamic
$J(x)=\Bigl(ie {{\partial}/{\partial x_{\mu}}}A_{\mu}(x)
+ie A_{\mu}(x){{\partial}/{\partial x_{\mu}}}-
e^2A_{\mu}(x)A^{\mu}(x) \Bigr)\Phi(x)$ which is generated by 
the usual 4D gauge transformation $q_{\mu}'=q_{\mu}-e{A}_{\mu}(q)$.
In the present 5D formulation (6.4) Lagrangian and the corresponding 
source operators are divided into two sets
of fields which are defined on the different 5D hyperboloids   
${q^2}\pm{q_5^2}=\pm M^2$ (see table 1). But in the limit 
$M\Longrightarrow 0$ or 
$M\Longrightarrow \infty$ one of the sets of the fields 
disappear and we obtain the conventional QED formulation.

In the same way we can construct the equation of motion for the
fermion field operator

$$\Bigl( i\gamma_{\mu}{ {\partial}\over { \partial x_{\mu}} } -m_{F}\Bigr)
\Psi(x)= J(x) \eqno(6.6a)$$ 
using the appropriate  5D equation 

$$\Bigl( i\gamma_{\mu}{ {\partial}\over { \partial x_{\mu}} } -m_{F}\Bigr)
{\widetilde\psi}_a(x,x_5)= 
{\widetilde j}_a(x,x_5)=e\gamma_{\mu}{\widetilde A_a}^{\mu}(x,x_5)
{\widetilde\psi}_a(x,x_5),\eqno(6.6b)$$

where $J(x)=j_{a=1}(x,t_5)+j_{a=2}(x,t_5)$  and
$\Psi(x)=\psi_{a=1}(x,t_5)+\psi_{a=2}(x,t_5)$. Here
$M$ is not fixed and 
${\widetilde\psi}_a(x,x_5)$ satisfies the
constraint
$\Bigl({i/ M}{{\partial}/{\partial
x_5}}-\eta_a\Bigr){\widetilde \psi}_a(x,x_5)={\widetilde
l}_a(x,x_5)$ and ${\widetilde j}_a(x,x_5)=(-1)^{a-1}M^2
\Bigl({i/ M}{{\partial}/{\partial
x_5}}+\eta_a\Bigr){\widetilde l}_a(x,x_5)$.

The geometrical realization of the scalar electrodynamics 
with conformal and gauge transformations in the curved 7D configuration space
was given by Chang and Chodos \cite{Chodos}.
This generalization of the Kaluza-Klein model provides a specific dynamical
mechanism for the breakdown of conformal symmetry and mass generation.
The resulting equations of motion are also defined in the projective
five-dimensional space and afterwards the corresponding 5D equations 
 were reduced into usual 4D equation of motion for a charged scalar fields.   
But the translation and inversion procedure 
in the momentum and in the coordinate 
space are independent for the quantum fields. Therefore 
one can extend Chang and Chodos model   
using the additional
 mechanisms of the mass generation and the conformal symmetry 
breaking coming from the considered conformal group of transformations
of quantum fields in the momentum space.
The goal of this extension is  to fix the scale parameter $M$ and to try
to reproduce the observed mass 
spectrum of scalar or pseudo-scalar particles.   
But this investigation is out of the 
scope of present paper.

\par
{\underline {\bf Gauge $SU(2)\times U(1)$
theory}} can by formulated in the
5D form with   
off $q^2\pm {q_5}^2=\pm M^2$ shell Lagrangian
following notations in \cite{ChL}

$${\widetilde {\cal L}}(x,x_5)=\sum_{a=1,2}
({\widetilde {\cal L}}_a)_V(x,x_5)+({\widetilde {\cal
L}}_a)_{sk}(x,x_5)+({\widetilde {\cal L}}_a)_F(x,x_5),
\eqno(6.7)$$
where $({\widetilde {\cal L}}_a)_V$ denotes the vector part of Lagrangian
with  Yang-Mills
 fields $({\widetilde A}_a)_{\mu}^{\alpha}(x,x_5)$ 
and Abelian fields $({\widetilde B}_a)_{\mu}(x,x_5)$

$$({\widetilde {\cal L}}_a)_V=
  -{1\over 4}(F_a)_{\mu\nu}^{\alpha}{(F_a)^{\mu\nu}}^{\alpha}
-{1\over 4}(G_a)_{\mu\nu}{(G_a)^{\mu\nu}}$$ $$ -M^2|{i\over
M}{{\partial}\over {\partial x_5}}({\widetilde
A}_a)^{\alpha}_{\mu}- ({\widetilde A}_a)^{\alpha}_{\mu}
-({\widetilde l}_a^A)^{\alpha}_{\mu}|^2 -M^2|{i\over
M}{{\partial}\over{\partial x_5}}({\widetilde B}_a)_{\mu}-
({\widetilde B}_a)_{\mu}- ({\widetilde l}_a^B)_{\mu}|^2,
\eqno(6.8a)$$ where
$$(F_a)_{\mu\nu}^{\alpha}=
{{\partial}\over{\partial{x^\mu}}}({\widetilde
A}_a)^{\alpha}_{\nu}-
{{\partial}\over{\partial{x^\nu}}}({\widetilde
A}_a)^{\alpha}_{\mu}+ g\varepsilon^{\alpha\beta\gamma}
({\widetilde A}_a)^{\beta}_{\mu} ({\widetilde
A}_a)^{\gamma}_{\nu},\eqno(6.8b)$$

$$(G_a)_{\mu\nu}=
{{\partial}\over{\partial{x^\mu}}}({\widetilde B}_a)_{\nu}-
{{\partial}\over{\partial{x^\nu}}}({\widetilde B}_a)_{\mu}.
\eqno(6.8c)$$

We can determine  $({\widetilde l}_a^A)^{\alpha}_{\mu}$,
$({\widetilde l}_a^B)_{\mu}$  via
 $({\widetilde j}_a^A)^{\alpha}_{\mu}$, $({\widetilde j}_a^B)_{\mu}$
in the same way as in eq.(6.3a).
The interacting parts of Lagrangian with  the gauge fields
$({\widetilde B}_a)_{\mu}$ and
$({\widetilde A}_a)^{\alpha}_{\mu}$ are included in the
fermion and in the scalar Lagrangians  $( {\widetilde
{\cal L}}_a)_F$ and $({\widetilde {\cal L}}_{a})_{sk}$.

The scalar part of Lagrangian  (6.7) is

$$({{\widetilde {\cal L}}_a})_{sk}=\Bigl(
{\cal D}_a^{\mu}{\widetilde \Phi}_a\Bigr)^{\dag} \Bigl({\cal
D}_{a\mu}{\widetilde \Phi}_a\Bigr)- M^2| {i\over
M}{{\partial}\over{\partial x_5}}({\widetilde \Phi}_a)-
{\widetilde \Phi}_a-{\widetilde l}^{\Phi}_a|^2  +
{({\widetilde{\cal L}}_a)}_{INT},\eqno(6.9a)$$
where $({\widetilde{\cal L}}_a)_{INT}$ contains the self-interaction 
term of the scalar particle and

$${\cal D}_{a\mu}{\widetilde \Phi}_a=
\Bigl({{\partial}\over{\partial{x^\mu}}}- ig {{\tau^{\alpha}}\over
2}({\widetilde A}_a)^{\alpha}_{\mu} - ig'({\widetilde
B}_a)_{\mu}\Bigr){\widetilde \Phi}_a .\eqno(6.9b)$$

Unlike  the standard $SU(2)\times U(1)$ theory, we start from the
massless  $\Phi_a(x,x_5)$ field and the Higgs mechanism we shall reproduce
using

$${\widetilde l}^{\Phi}_{a}(x,x_5)=
-{f\over M}{\widetilde \Phi}_{a}(x,x_5) \Bigl({\widetilde
\Phi}_{a}^{\ast}(x,x_5) {\widetilde
\Phi}_{a}(x,x_5)\Bigr)^{1/2}.\eqno(6.10)$$

In the unitary gauge

$$
{\widetilde \Phi}_{a}(x,x_5)={\cal U}\Bigl({\widetilde
 \zeta}_a(x,x_5)\Bigr)
\left( \begin{array}{c} 0  \\
 {\widetilde \phi}_a(x,x_5)\end{array} \right), \eqno(6.11a)$$
where

$${\cal U}\Bigl({\widetilde \zeta}_a(x,x_5)\Bigr)=
exp\Bigl(i{\widetilde \zeta}_a(x,x_5)/v\Bigr). \eqno(6.11b)$$

We shall assume, that ${\widetilde \zeta}_a$ is independent on $x_5$

$${{\partial}\over{\partial{x_5}}}
{\widetilde \zeta}_a(x,x_5)=0.\eqno(6.12)$$

Afterwards ${\widetilde l}_a$ (6.10) takes a form
$${\widetilde l}^{\Phi}_{a}(x,x_5)=-{f\over M}{\cal U}\Bigl({\widetilde
\zeta}_a(x,x_5)\Bigr)
\left( \begin{array}{c} 0  \\
{{\widetilde \phi}_a}(x,x_5) \end{array}
\right)\sqrt{\Bigl({{\widetilde \phi}_a}(x,x_5)\Bigr)^2},
\eqno(6.13)$$
and for ${i/ M}{{\partial}/{\partial
x_5}}{\widetilde \phi}_a$ we get

$${i\over M}{{\partial}\over{\partial x_5}}({\widetilde \phi}_a)
={\widetilde \phi}_a-{f\over M}{\widetilde
\phi}_a\sqrt{({\widetilde \phi}_a)^2}.\eqno(6.14)$$
Next for ${\widetilde j}_a$ and for
$({\widetilde {\cal L}}_a)_{INT}$ we have

$${\widetilde j}_{a}=(-1)^{a-1}{\cal U}({\widetilde \zeta}_a)M^2
\left( \begin{array}{c} 0  \\
1 \end{array} \right) \Bigl( -3f M{\widetilde
\phi}_{a}\sqrt{({\widetilde \phi}_a)^2}+2f^2 {\widetilde
\phi}_a^3\Bigr) \eqno(6.15)$$

and

$$({\widetilde {\cal L}}_a)_{INT}=(-1)^{a-1}\Bigl( -fM{\widetilde
\phi}_a^2\sqrt{({\widetilde \phi}_a)^2}
 +{{f^2}\over 2}{\widetilde \phi}_a^4\Bigr)\eqno(6.16)$$

For the positive $f$  Lagrangian  ${\widetilde {\cal L}}_{a=1\equiv
  inr}$ is similar to the self-interaction potential  
$$V(\Phi)=-\mu^2\Phi^2+\lambda\Phi^4.\eqno(6.17)$$
In particular ${\widetilde {\cal L}}_{1}$ has zero at
${\widetilde \phi}_1=0$ and ${\widetilde \phi}_1= \pm {{2M}/ f}$
and  ${\widetilde {\cal L}}_{1}$ has minima at ${\widetilde
\phi}_1=\pm {{3M}/ {2f}}$. It is important to note, that
$\biggl[{\widetilde {\cal L}}_{a=2}\biggr]_{INT}=-
\biggl[{\widetilde {\cal L}}_{a=1}\biggr]_{INT}$.
Therefore in the mass term of Lagrangian 
the real positive $m^2$ can arise only   
in ${\widetilde {\cal L}}_{1}$. In order to taken into account
the spontaneous symmetry breaking mechanism we introduce

$$<0|\phi_{a}(x,x_5)|0>={1\over{2}}
\left( \begin{array}{c} 0  \\
{\sl v}_a\end{array} \right),\eqno(6.18)$$ where
 ${\sl v}_{a=1}={\sl v}$ and ${\sl v}_{a=2}=0$.
Afterwards we get

$${\widetilde \phi}_{1}(x,x_5)={1\over2 }
\left( \begin{array}{c} 0  \\
{\sl v}+{\widetilde \phi_1}'(x,x_5)\end{array} \right);\ \ \ \ \
 {\widetilde \phi}_{2}(x,x_5)={1\over{2}}
\left( \begin{array}{c} 0  \\
{\widetilde \phi_2}'(x,x_5)\end{array} \right) .\eqno(6.19)$$
Therefore Lagrangian  (6.16) takes the form
$$({\widetilde {\cal L}}_a)_{INT}=(-1)^{a-1}\Bigl(
-{{fM}\over 4}({{\widetilde \phi}_{a}}'+{\sl v}_a )^2
        \sqrt{({{\widetilde \phi}_{a}}'+{\sl v}_a )^2} + {{f^2}\over
16}({\widetilde \phi}_{a}'+{\sl v}_a
 )^{4}\Bigr). \eqno(6.20)$$

Thus instead of the mass term in
the usual 4D self interacting potential (6.17)
(see for instance ch. 8 and 11 of \cite{ChL})
in the 5D Lagrangian  (6.20) arise the following mass terms 
 $$\mu^2 \Bigl({1\over{\sqrt{2} }}(\Phi'+v)\Bigr)^2\Longrightarrow
 {{fM}\over 4} \Bigl({{\widetilde \phi}'}_{1}+{\sl v}\Bigr)^{2}
        \sqrt{\Bigl({{\widetilde \phi}'}_{1}+{\sl v}\Bigr)^{2}}
-{{fM}\over 4}      ({{\widetilde \phi}'}_{2} )^2
         \sqrt{({{{\widetilde \phi}'}_{2}} )^2 } ,\eqno(6.21)$$
which determine the mass spectrum.
\footnotemark.

The effective Lagrangian (6.21) has minima at
 ${\widetilde \phi}_1=\pm {{3M}/ {f}}$. Therefore
  ${\widetilde {\cal L}}_{a=1}$ does not contain
the linear terms  in ${\widetilde \phi}_1$,
i.e. ${\widetilde j}_1$ does not include the  constant terms.
If we put $v=\sqrt{2}{\sl v}=3\sqrt{2}M/f$,
then  all expressions for the  fermion masses  in the considered model
were reproduced, i.e.
 $m_{k}=f_kv/\sqrt{2}$, $k=el,u,d$
and the  $W,Z$ meson masses remain be the same 
$m_Z=m_W/cos{\theta_W}$. But, for the Higgs boson mass
term in Lagrangian (6.21) we get $m_{higgs}^2=9/8M^2$.
Thus the principal difference between the suggested 5D formulation
and the standard
$SU(2)\times U(1)$ gauge field theory
is generated by the symmetry breaking terms (6.21).
In the present model the scale parameter $M$  is
determined via the mass of the Higgs boson and it 
indicates  the border  of regions $q^2=0,\pm M^2$
 where the interaction of the scalar fields change the sign.

\par
\footnotetext{
According to the definition of action (4.2) and field operators (3.11a,b)
present formulation is four-dimensional in the momentum space.
Therefore Lagrangians (6.16) and (6.20) remain to be renormalizable.}
\vspace{0.5cm}

\centerline{\bf{7.\ Summary and outlook}} \vspace{0.5cm}
\par

In this paper we have considered 
 conformal transformations of the interacted quantum fields
in the momentum space. 
In the quantum field theory translations and inversions in the coordinate 
and in the momentum spaces are independent each from other. 
Therefore we can separate two different
$O(2,4)$ groups of the conformal transformations: one for the well known 
conformal transformations in the coordinate space and the other  
in the momentum space.  This offers an additional possibilities by 
applications of the conformal groups in the quantum field theory. 
Here we have applied    
these conformal transformations of quantum fields in the momentum space
to the  chiral symmetry breaking models  with and without gauge fields.

Unlike to the other papers devoted to the conformal transformations,
we operate from the beginning with the interacting fields with mass
and other scale parameters. 
Therefore any  mass generation 
procedure can be incorporated in the considered scheme.
The key point of the present formulation is the invariance of the 
6D cone $\kappa_{\mu}\kappa^{\mu}+\kappa_5^2-\kappa_6^2=0$
under the conformal transformations.
Therefore the 5D forms  $q^2\pm q_5^2=\pm M^2$ (2.2a,b), 
arising after  projection
of this 6D cone into 5D momentum space with
$q_{\mu}={ {\kappa_{\mu}}/{\kappa_{+} }};\ 
\kappa_{\pm}=(\kappa_{5}\pm\kappa_{6})/M;\ 
\mu=0,1,2,3$ (1.3), are also invariant.  
This invariance of 5D forms was obviously preserved  by derivation 
of the 5D equation of motion 
$\Bigl({ {\partial^2}/ { \partial x_{\mu} \partial x^{\mu}}} +m^2\Bigr)
\varphi_{a}(x,x_5)=j_{a}(x,x_5)$,  
$(a=1,2)$ and derivation of the  
boundary condition for the fifth coordinate  
${i/ M}{ {\partial}/ { \partial x_5}}\varphi_a (x,x_5) =
\eta_a\varphi_a(x,x_5)+l_a(x,x_5)$ (3.4), where
$j_{a}(x,x_5)=(-1)^{(a-1)}
M^2\Bigl({i/ M}{ {\partial}/ { \partial x_5}}+\eta_{a}\Bigr)
l_{a}(x,x_5)$.
In addition, operators $\varphi_{a}(x,x_5)$, $l_{a}(x,x_5)$,$j_{a}(x,x_5)$  
are embedded in the different 5D space with two different invariant forms 
$q^2\pm q_5^2=\pm M^2$. Therefore they satisfy boundary conditions 
$\biggl({{\partial^2}/{\partial x^{\mu}\partial x_{\mu}}}\pm
{{\partial^2}/{\partial x^5\partial x_5}}\pm M^2\biggr){\cal O}_a(x,x_5)=0,$ 
where  ${\cal O}_a=\varphi_{a},l_{a},j_{a}$ (see (3.3a,b) and (3.10a,b,c)).
The 4D field operator $\Phi(x)$ is determined via  $\varphi_a(x,x_5)$ as  
$\Phi(x)=\Phi(x,x_5=t_5)=\varphi_{1}(x,x_5=t_5)+\varphi_{2}(x,x_5=t_5)$,
 where $t_5=\sqrt{x_o^2-{\bf x}^2}$ or $t_5=0$. Therefore from the 
above set of the 5D 
equations we obtain the ordinary 4D equation of motion
$\biggl({{\partial^2}/{\partial x^{\mu}\partial x_{\mu}}}+
m^2\biggr)\Phi(x)=J(x)$ with 
$J(x)=J_{1}(x,x_5=t_5)+J_{2}(x,x_5=t_5)$.

Separation of the boundary conditions over the fifth coordinates 
${i/ M}{ {\partial}/ { \partial x_5}}\varphi_a (x,x_5) =
\eta_a\varphi_a(x,x_5)+l_a(x,x_5)$ with
$j_{a}(x,x_5)=(-1)^{(a-1)}
M^2\Bigl({i/ M}{ {\partial}/ { \partial x_5}}+\eta_{a}\Bigr)
l_{a}(x,x_5)$ makes it possible to interpret $l_a(x,x_5)$ as 
a translation operator of the fifth coordinate. But unlike to the 
other five-dimensional formulations \cite{Fanchi,Land,Kubo,Green},
these one dimensional equation can not be interpreted as the evolution  
equations, because $l_a$ are the determined as the second variation from the 
Lagrangians.

The considered scheme of 5D projections of the 6D conformal invariant
forms can be realized also in the coordinate space.
Thus the 6D invariant form $\xi_A\xi^A=0$ with
$x_{\mu}=\xi_{\mu}{\ell}/(\xi_5+\xi_6)$ and
$x^{\mu}x_{\mu}=-\ell^2(\xi_5-\xi_6)/(\xi_5+\xi_6)$ can be
projected into two 5D hyperboloids
$x^{\mu}x_{\mu}\pm x_5^2=\pm\ell^2$
with $x_5^2=2\xi_5(\ or\ \xi_6)\ell^2/(\xi_5+\xi_6)$.
This leads to separation of the internal  and the external 
5D regions with the boundary values at $x^2=0,\pm \ell^2$.
Besides, in this approach arise an additional boundary condition for the
fifth coordinate for construction
of the 4D field operator
$\Phi(x)=\varphi_{inr}(x,x_5=t)+\varphi_{ext}(x,x_5=t)$
at $t=0\ or\ t=\sqrt{x^2}$. The advantage of this scheme is that
there is introduced a new scale  parameter $\ell$ which can be
used by description of observables.

Considered above field operators are defined in the 
5D space with the invariant forms
$q^2+ q_5^2=M^2$ or $q^2- q_5^2=- M^2$ correspondingly. 
These  areas are connected with inversion
$q'_{\mu}=-M^2 q_{\mu}/q^{2}$. Present formulation contains 
 an additional mass parameter $M$
which indicates the scale of the broken conformal symmetry.
Moreover
the interaction 5D Lagrangians and the 
corresponding source operators change their sign at the border 
$q^2=0$ and $q^2=\pm M^2$.
We have shown, that the simple choice of $l_a$ (5.10) and
 (6.10) leads to the  5D generalization of the non-linear $\sigma$-model
and of the standard $SU(2)\times U(1)$ theory.
Corresponding 5D Lagrangians contains the specific mechanism of the
breakdown of the chiral symmetry and  
allows us to determine $M$ through the mass of pion or Higgs boson.
Unlike to the usual $\sigma$-model,
the present 5D Lagrangian contains the 
standard weak chiral symmetry breaking term 
$\lambda\sigma$ together with other terms which
broke the chiral symmetry more strongly. 
For the gauge theories without the chiral symmetry like QED,
where $M$ is not fixed, it was shown, 
that in the limit $M\Longrightarrow \infty$ or $M\Longrightarrow 0$ we obtain 
the conventional 4D formulation of QED.

\vspace{0.5cm}

I am sincerely grateful to V.G.Kadyshevsky for numerous constructive 
and fruitful discussions. 



\end{document}